\documentclass{article}
\usepackage{emulateapj}
\usepackage{apjfonts}
\usepackage[american]{babel}

\newcommand{\kms}{\ensuremath{\mathrm{km\ s^{-1}}}}
\newcommand{\fe}{\ensuremath{\langle\mathrm{Fe}\rangle}}
\newcommand{\mgb}{\ensuremath{\mathrm{Mg}\,b}}
\newcommand{\mg}{\ensuremath{\mathrm{Mg}}}

\newcommand{\mgtwo}{\ensuremath{\mathrm{Mg}_2}}
\newcommand{\umb}{\ensuremath{(U-B)}}
\newcommand{\umv}{\ensuremath{(U-V)}}
\newcommand{\bmv}{\ensuremath{(B-V)}}
\newcommand{\hbeta}{\ensuremath{\mathrm{H}\beta}}
\newcommand{\logt}{\ensuremath{\log t}}
\newcommand{\z}{\ensuremath{\mathrm{[Z/H]}}}
\newcommand{\zp}{\ensuremath{\mathrm{Z}}}
\newcommand{\feh}{\ensuremath{\mathrm{[Fe/H]}}}
\newcommand{\mgh}{\ensuremath{\mathrm{[Mg/H]}}}
\newcommand{\enh}{\ensuremath{\mathrm{[E/Fe]}}}
\newcommand{\eh}{\ensuremath{\mathrm{[E/H]}}}
\newcommand{\reo}[1]{\ensuremath{r_{e}/#1}}

\newcommand{\tzas}{$t$--\z--\enh--$\sigma$}
\newcommand{\vos}{\ensuremath{(v/\sigma_0)^{\ast}}}

\def\plotonet#1{\centering \leavevmode
    \epsfxsize=0.95\textwidth \epsfbox{#1}}

\submitted{Accepted for publication in the Astronomical Journal}
 
\lefthead{TRAGER ET AL.}
\righthead{STELLAR POPULATIONS OF EARLY-TYPE GALAXIES. II. CONTROLLING
PARAMETERS}
 
\begin{document}

\title{The Stellar Population Histories of Early-Type Galaxies.
II. Controlling Parameters of the Stellar Populations}

\author{S. C. Trager\altaffilmark{1,2}}
\affil{The Observatories of the Carnegie Institution of Washington,
813 Santa Barbara St., Pasadena, CA 91101; sctrager{\@@}ociw.edu}
\authoraddr{813 Santa Barbara St., Pasadena, CA 91101;
sctrager{\@@}ociw.edu}
\altaffiltext{1}{Carnegie Starr Fellow}
\altaffiltext{2}{Hubble Fellow}

\author{S. M. Faber}
\affil{UCO/Lick Observatory and Board of Studies in Astronomy and
Astrophysics, University of California, Santa Cruz, Santa Cruz, CA
95064; faber{\@@}ucolick.org}
\authoraddr{University of California, Santa Cruz, CA 95064;
faber{\@@}ucolick.org}

\author{Guy Worthey}
\affil{Department of Physics and Astronomy, St.~Ambrose University,
Davenport, IA 52803-2829; gworthey{\@@}saunix.sau.edu}
\authoraddr{Davenport, IA 52803-2829; gworthey{\@@}saunix.sau.edu}

\author{J. Jes\'us Gonz\'alez}
\affil{Instituto de Astronom{\'\i}a---UNAM, Apdo Postal 70-264,
M\'exico D.F., Mexico; jesus{\@@}astroscu.unam.mx}
\authoraddr{Apdo Postal 70-264, M\'exico D.F., Mexico;
jesus{\@@}astroscu.unam.mx}

\begin{abstract}

This paper analyzes single-stellar-population (SSP) equivalent
parameters for 50 local elliptical galaxies as a function of their
structural parameters.  The galaxy sample is drawn from the
high-quality spectroscopic surveys of Gonz\'alez (1993) and Kuntschner
(1999).  The basic data are central values of SSP-equivalent ages,
$t$, metallicities, \z, and ``enhancement'' ratios, \enh, derived in
Paper I, together with global structural parameters including velocity
dispersions, radii, surface brightnesses, masses, and luminosities.

The galaxies fill a two-dimensional plane in the four-dimensional
space of \z, $\logt$, $\log\sigma$, and \enh.  SSP age, $t$ and
velocity dispersion, $\sigma$, can be taken as the two independent
parameters that specify a galaxy's location in this ``hyperplane.''
The hyperplane can be decomposed into two sub-relations: (1) a
``\zp-plane,'' in which \z\ is a linear function of $\log\sigma$ and
$\logt$; and (2) a relation between \enh\ and $\sigma$ in which \enh\
is larger in high-$\sigma$ galaxies.  Velocity dispersion is the only
structural parameter that is found to modulate the stellar
populations; adding other structural variables such as $I_e$ or $r_e$
does not predict \z\ or \enh\ more accurately.

Cluster and field ellipticals follow the same hyperplane, but their
$(\sigma,t)$ distributions within it differ.  Most Fornax and Virgo
cluster galaxies are old, with a only a small sprinkling of galaxies
to younger ages.  The field ellipticals span a larger range in SSP
age, with a tendency for lower-$\sigma$ galaxies to be younger.  The
present sample thus suggests that the distribution of local
ellipticals in the $(\sigma,t)$ plane may depend on environment.
Since the $(\sigma,t)$ distribution affects all two-dimensional
projections involving SSP parameters, many of the familiar scaling
laws attributed to ellipticals may also depend on environment.  Some
evidence for this is seen in the current sample.  For example, only
Fornax ellipticals show the classic mass-metallicity relation, whereas
other sub-samples do not.

The tight Mg--$\sigma$ relations of these ellipticals can be
understood as two-dimensional projections of the metallicity
hyperplane showing it edge on.  At fixed $\sigma$, young age tends to
be offset by high \z, preserving Mg nearly constant.  The tightness of
the Mg--$\sigma$ relations does not necessarily imply a narrow range
of ages at fixed $\sigma$.

Although SSP parameters are heavily weighted by young stars, modeling
them still places tight constraints on the total star formation
history of elliptical galaxies.  The relation between \enh\ and
$\sigma$ is consistent with a higher effective yield of Type II SNe
elements at higher $\sigma$.  This might occur if the IMF is enhanced
in massive stars at high $\sigma$, or if more SNe II-enriched gas is
retained by deeper galactic potential wells.  Either way, modulating
Type II yields vs. $\sigma$ seems to fit the data better than
modulating Type Ia yields.

The \zp-plane is harder to explain and may be a powerful clue to star
formation in elliptical galaxies if it proves to be general.  Present
data favor a ``frosting'' model in which low apparent SSP ages are
produced by adding a small frosting of younger stars to an older
``base'' population (assuming no change in $\sigma$).  If the frosting
abundances are close to or slightly greater than the base population,
simple two-component models run along lines of constant $\sigma$ in
the \zp-plane, as required. This favors star formation from well-mixed
pre-enriched gas rather than unmixed low-metallicity gas from an
accreted object.

\end{abstract}


\section{Introduction}

The star formation histories of elliptical galaxies, once thought to
be very simple---old and metal-rich (\cite{Baade63})---have come under
increasing scrutiny in the last three decades (e.g., \cite{ST71};
\cite{Faber73}, 1977; \cite{O'Connell76}, 1980; \cite{Pickles85};
\cite{P89}; \cite{SS92}; \cite{G93}; \cite{Worthey94}; \cite{Lee94};
\cite{Renzini95}, 1998; \cite{TCB98}; \cite{K98}; \cite{J99}).
Currently there are two basic models for elliptical galaxy formation:
hierarchical clustering of small objects into larger galaxy-sized
units with accompanying star formation over time (e.g., \cite{BFPR84};
\cite{KWG93}), versus monolithic collapse and star formation in a
nearly coeval single early burst (e.g., \cite{ELS62}; \cite{Larson74};
\cite{AY87}).  Measurements of the spectral energy distributions
(SEDs) and spectral features of elliptical galaxies can provide a test
of these scenarios.  For example, evidence for substantial
intermediate-age stellar populations (between 1 and 10 Gyr) might
favor hierarchical models, which more naturally have extended star
formation over time.  A goal of the present series is to assess the
evidence for such intermediate-age populations.

The first paper of this series (\cite{Paper1}, hereafter Paper I) used
Lick absorption-line strengths for a sample of local elliptical
galaxies observed by Gonz{\'a}lez (1993, hereafter G93) to derive
single-stellar-population (SSP) equivalent parameters $t$ (age), \z\
(metallicity), and \enh\ (``enhancement ratio,'' see below).
Single-burst model line-strengths by Worthey (1994, hereafter W94)
were corrected for the effect of non-solar abundance ratios using
theoretical spectral calculations by Tripicco \& Bell (1995, hereafter
TB95).  The resultant SSP ages cover a range of 1 to 18 Gyr (including
observational errors), while the ranges in \z\ and \enh\ are fairly
narrow.  These parameters, particularly the ages, are based on the
assumption that \hbeta\ faithfully traces the mean temperature of the
main-sequence turnoff and is not seriously affected by other hot
stellar populations.  Evidence supporting this assumption was
presented in Paper I.

In deriving single-burst SSP parameters for elliptical galaxies, we do
not mean to imply that their star formation histories were actually
single bursts.  In fact, our favored ``frosting'' model
(Section~\ref{sec:disc_plane}) involves adding a minority of young
stars to an older base population.  Our use of SSP parameters is
simply a convenient way of condensing all the presently measured line
strength data into just three numbers: light-weighted age, \z, and
\enh.  For the moment, that is all the observations allow.  It is our
hope that SSP parameters will be adopted by those who model the full
evolutionary history of elliptical galaxies (e.g., \cite{AY87};
\cite{Vazdekis96}; \cite{TCBMP98}) and that they will serve as as a
convenient meeting ground between models and data.  We show below
that, even though SSP parameters are heavily influenced by the light
of any young stars that may be present, modeling them still places
important constraints on the total history of star formation in
ellipticals.

This paper explores the central stellar populations of a sample of
local elliptical galaxies and develops correlations among them and
with parent-galaxy structural parameters.  Many previous works have
studied such correlations, but most have focused on \emph{raw} line
strengths.  Only three other studies, to our knowledge, have measured
ages (using Balmer lines) and developed correlations based on
underlying stellar populations.  Tantalo, Chiosi \& Bressan (1998)
studied the G93 galaxies using models based on the ``Padua''
isochrones.  Their correction for non-solar abundance ratios was
approximate, however, leading to systematic errors in derived age, \z,
and \enh\ (Paper I).  Kuntschner (1999) studied ellipticals in Fornax
using high-quality data, which we add to our sample here.  He found
that Fornax ellipticals were mainly old, and also discovered a strong
relation between \enh\ and $\sigma$, which we confirm.  J{\o}rgensen
(1999) studied Coma ellipticals using line-strength models by Vazdekis
et al.~(1996).  Her conclusions foreshadow ours in many respects, but
some seem in retrospect to be the product of observational errors.
All three of these papers are discussed in
Section~\ref{sec:sum_compare}.

The outline of this paper is as follows.  A brief review of the G93
and Kuntschner (1999) samples, line-strength data, SSP-equivalent
stellar population parameters, and structural parameters is given in
Section~\ref{sec:data}.  Section~\ref{sec:results} presents the sample
distribution in the four-dimensional space spanned by \z, $\logt$,
$\log\sigma$, and \enh; this proves to be a highly flattened,
two-dimensional ``hyperplane'' that in turn consists of two
sub-relations, a ``\zp-plane,'' plus a linear \enh--$\sigma$ relation.
Section~\ref{sec:other_views} shows how projections of this hyperplane
depend on the distribution of points within it, and thus how the
appearance of two-dimensional scaling laws can vary depending on this
distribution. Section~\ref{sec:scaling_laws} illustrates these effects
using two classic scaling laws---the mass-metallicity relation and the
Mg--$\sigma$ relation.  Possible evidence for environmental variation
in the former is presented.  Sections~\ref{sec:disc_sz} and
\ref{sec:disc_plane} investigate the origins of the hyperplane.  The
\zp-plane in particular appears difficult to explain and, if it proves
general, will place very tight constraints on the history of star
formation in local elliptical galaxies.  Section~\ref{sec:conc}
summarizes our findings and conclusions.

\section{Data and derived parameters}\label{sec:data}

This section briefly describes the G93 and Fornax samples, the
Lick/IDS line-strength system, the models used to transform line
strengths into SSP-equivalent parameters, and final population
parameters for the central (\reo{8}) aperture observations of G93 and
Fornax ellipticals.  A complete description of the data and their
transformation into stellar population parameters was given in Paper
I.  Structural parameters drawn from the literature for these galaxies
are also given.

\subsection{Sample}

Trager (1997) showed that deriving SSP parameters from Balmer and
metal lines requires line-strength data of very high quality, with
errors preferably $<0.1$\AA.  Only three published samples approach
this level of accuracy: Gonz{\'a}lez (1993), Kuntschner (1998), and
Fisher, Franx, and Illingworth (1995).  The original G93 sample
consists of 41 early-type galaxies, of which 40 are included in the
present study (NGC 4278 is discarded because of its strong emission).
All G93 galaxies used here have been classified as elliptical (or
compact elliptical) in the RC3 (\cite{RC3}) or the RSA (\cite{RSA})
and Carnegie Atlas (\cite{CA}), except for NGC 507 and NGC 6703, both
classified as SA0 in the RC3 but not included in the RSA or Carnegie
Atlas, and NGC 224, the bulge of the Sb galaxy Messier 31.

The environmental distribution of the G93 sample is skewed toward
relatively low-density environments.  As discussed in Paper I, most of
the galaxies are in small groups of varying richness, many are
relatively isolated, and six are members of the Virgo Cluster.  Only
one is in a rich cluster (NGC 547 in Abell 194).  Environmental
effects on the stellar populations of ellipticals are discussed in
Sections~\ref{sec:other_views} and \ref{sec:scaling_laws} below.

The G93 sample is augmented here with data from Kuntschner (1999,
hereafter K98; cf.\ \cite{KD97}) on early-type galaxies in the Fornax
cluster.  These data have been carefully transformed to the Lick
line-strength system.  Eleven of the 22 galaxies in K98 are
ellipticals.  SSP parameters have been derived for them following the
method below, after correcting the central line strengths (Table 3.4
of K98) to the \reo{8}\ aperture using the gradients presented in
Table 7.2 of K98.

The high-accuracy elliptical galaxy sub-sample of Fisher et al. (1995)
repeats galaxies in G93 and agrees well with it.  These data have
therefore not been used here.

\subsection{Ages, metallicities, and enhancement ratios}

\begin{figure*}
\plotonet{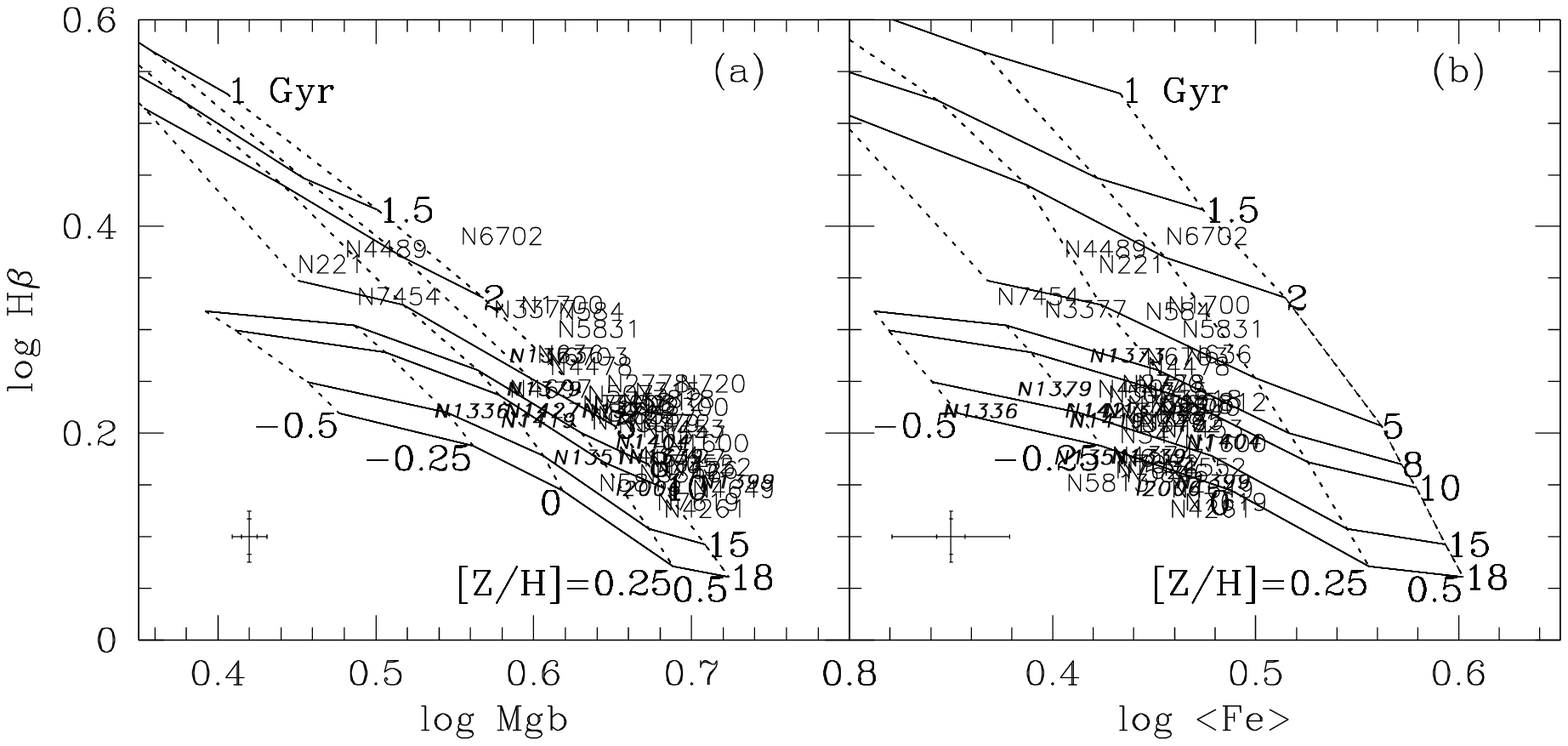}
\caption{Line strengths of G93 (roman type; smaller error bars) and
Fornax ellipticals (slanted bold type; larger error bars) through the
central \reo{8}\ aperture.  Solar-ratio model grids from Worthey
(1994) have been superimposed: solid lines are contours of constant
age (from top, 1, 1.5, 2, 5, 8, 10, 15, 18 Gyr), and dotted lines are
contours of constant \z\ (from left, $\z=-0.5$, $-0.25$, 0.0, +0.25,
+0.5 dex, except at ages younger than 8 Gyr, where from left
$\z=-0.225$, 0.0, +0.25, +0.5 dex). (a) \mgb\ and \hbeta\ line
strengths. (b) \fe\ and \hbeta\ line strengths.  Differences in the
ages and metallicities inferred from the two diagrams result from the
non-solar abundance ratios of giant elliptical galaxies.  Our
procedure corrects for this, and in so doing derives the non-solar
abundance ratio, \enh.\label{fig:re8m}}
\end{figure*}

Paper I describes our technique for inverting line strengths to
determine SSP parameters.  Ages, metallicities, and enhancement ratios
of old stellar populations are determined by comparing observed
absorption-line strengths to the \emph{single-burst stellar population
(SSP)} models of W94, which depend on metallicity and age.  The
line-strengths of the Worthey models correspond to solar abundance
ratios; these have been corrected for non-solar abundance ratios as
described in Paper I using the theoretical spectral calculations of
TB95, who tabulated the response of the Lick/IDS indices to changes in
the abundance ratios of important elements.  SSP-equivalent $t$, \z,
and \enh\ are derived for each galaxy by searching a finely-spaced
grid of points in $(\hbeta,\mgb,\fe)$ space.  Central line strengths
corrected to the \reo{8}\ aperture are presented in
Figure~\ref{fig:re8m} for the G93 and K98 samples.

\begin{deluxetable}{lrrrrr}
\tablecaption{Central ages, metallicities and enhancement ratios
through \reo{8} aperture (enrichment model 4)\label{tbl:re8tza}}
\tablewidth{0pt}
\tablehead{\colhead{Name}&\colhead{Age (Gyr)}&
\colhead{\z}&\colhead{\enh}&\colhead{\feh}&\colhead{\eh}}
\startdata
\sidehead{G93 ellipticals:}
NGC 221&$3.0\pm0.7$&$0.00\pm0.05$&$-0.08\pm0.01$&$0.07\pm0.05$&$-0.01\pm0.05$\nl
NGC 224&$6.0\pm1.6$&$0.39\pm0.05$&$0.19\pm0.02$&$0.21\pm0.05$&$0.40\pm0.05$\nl
NGC 315&$5.4\pm1.5$&$0.34\pm0.06$&$0.25\pm0.02$&$0.11\pm0.06$&$0.36\pm0.06$\nl
NGC 507&$7.4\pm2.8$&$0.19\pm0.07$&$0.20\pm0.03$&$0.00\pm0.08$&$0.20\pm0.07$\nl
NGC 547&$8.9\pm2.4$&$0.24\pm0.05$&$0.26\pm0.02$&$-0.00\pm0.05$&$0.26\pm0.05$\nl
NGC 584&$2.5\pm0.3$&$0.49\pm0.03$&$0.22\pm0.01$&$0.29\pm0.03$&$0.51\pm0.03$\nl
NGC 636&$4.1\pm0.7$&$0.34\pm0.07$&$0.11\pm0.02$&$0.24\pm0.07$&$0.35\pm0.07$\nl
NGC 720&$4.5\pm2.3$&$0.46\pm0.17$&$0.33\pm0.04$&$0.15\pm0.17$&$0.48\pm0.17$\nl
NGC 821&$7.5\pm1.2$&$0.23\pm0.03$&$0.15\pm0.01$&$0.09\pm0.03$&$0.24\pm0.03$\nl
NGC 1453&$7.6\pm1.9$&$0.32\pm0.06$&$0.22\pm0.02$&$0.12\pm0.06$&$0.34\pm0.06$\nl
NGC 1600&$8.1\pm2.2$&$0.37\pm0.06$&$0.23\pm0.02$&$0.16\pm0.06$&$0.39\pm0.06$\nl
NGC 1700&$2.3\pm0.3$&$0.50\pm0.03$&$0.16\pm0.01$&$0.35\pm0.03$&$0.51\pm0.03$\nl
NGC 2300&$5.9\pm1.5$&$0.38\pm0.05$&$0.25\pm0.02$&$0.15\pm0.05$&$0.40\pm0.05$\nl
NGC 2778&$5.4\pm1.8$&$0.30\pm0.09$&$0.23\pm0.03$&$0.09\pm0.09$&$0.32\pm0.09$\nl
NGC 3377&$3.7\pm0.8$&$0.20\pm0.06$&$0.20\pm0.02$&$0.01\pm0.06$&$0.21\pm0.06$\nl
NGC 3379&$8.6\pm1.4$&$0.22\pm0.03$&$0.21\pm0.01$&$0.02\pm0.03$&$0.23\pm0.03$\nl
NGC 3608&$6.9\pm1.5$&$0.26\pm0.05$&$0.17\pm0.02$&$0.10\pm0.05$&$0.27\pm0.05$\nl
NGC 3818&$5.6\pm1.8$&$0.37\pm0.08$&$0.23\pm0.03$&$0.16\pm0.08$&$0.39\pm0.08$\nl
NGC 4261&$15.5\pm3.3$&$0.19\pm0.04$&$0.20\pm0.01$&$0.00\pm0.04$&$0.20\pm0.04$\nl
NGC 4374&$12.2\pm2.2$&$0.13\pm0.03$&$0.21\pm0.01$&$-0.07\pm0.03$&$0.14\pm0.03$\nl
NGC 4472&$7.9\pm1.7$&$0.26\pm0.05$&$0.21\pm0.02$&$0.06\pm0.05$&$0.27\pm0.05$\nl
NGC 4478&$4.6\pm2.3$&$0.30\pm0.10$&$0.15\pm0.03$&$0.16\pm0.10$&$0.31\pm0.10$\nl
NGC 4489&$2.5\pm0.4$&$0.14\pm0.06$&$0.03\pm0.02$&$0.11\pm0.06$&$0.14\pm0.06$\nl
NGC 4552&$10.5\pm1.2$&$0.28\pm0.04$&$0.23\pm0.01$&$0.07\pm0.04$&$0.30\pm0.04$\nl
NGC 4649&$11.7\pm1.5$&$0.29\pm0.04$&$0.25\pm0.01$&$0.06\pm0.04$&$0.31\pm0.04$\nl
NGC 4697&$8.9\pm1.9$&$0.06\pm0.06$&$0.10\pm0.02$&$-0.03\pm0.06$&$0.07\pm0.06$\nl
NGC 5638&$8.3\pm1.4$&$0.20\pm0.03$&$0.19\pm0.01$&$0.02\pm0.03$&$0.21\pm0.03$\nl
NGC 5812&$5.3\pm1.1$&$0.39\pm0.04$&$0.20\pm0.01$&$0.20\pm0.04$&$0.40\pm0.04$\nl
NGC 5813&$18.3\pm2.3$&$-0.03\pm0.03$&$0.21\pm0.01$&$-0.23\pm0.03$&$-0.02\pm0.03$\nl
NGC 5831&$2.6\pm0.3$&$0.54\pm0.03$&$0.19\pm0.01$&$0.36\pm0.03$&$0.55\pm0.03$\nl
NGC 5846&$13.5\pm3.3$&$0.15\pm0.05$&$0.22\pm0.02$&$-0.05\pm0.05$&$0.17\pm0.05$\nl
NGC 6127&$11.6\pm2.2$&$0.18\pm0.04$&$0.23\pm0.02$&$-0.03\pm0.04$&$0.20\pm0.04$\nl
NGC 6702&$1.5\pm0.1$&$0.70\pm0.07$&$0.15\pm0.03$&$0.56\pm0.08$&$0.71\pm0.07$\nl
NGC 6703&$4.3\pm0.7$&$0.32\pm0.06$&$0.15\pm0.02$&$0.18\pm0.06$&$0.33\pm0.06$\nl
NGC 7052&$12.5\pm3.1$&$0.17\pm0.05$&$0.24\pm0.02$&$-0.05\pm0.05$&$0.19\pm0.05$\nl
NGC 7454&$5.0\pm1.0$&$-0.06\pm0.04$&$0.06\pm0.02$&$-0.12\pm0.04$&$-0.06\pm0.04$\nl
NGC 7562&$7.6\pm1.6$&$0.21\pm0.04$&$0.17\pm0.01$&$0.05\pm0.04$&$0.22\pm0.04$\nl
NGC 7619&$14.4\pm2.2$&$0.21\pm0.03$&$0.18\pm0.01$&$0.04\pm0.03$&$0.22\pm0.03$\nl
NGC 7626&$12.8\pm2.4$&$0.17\pm0.03$&$0.25\pm0.01$&$-0.06\pm0.03$&$0.19\pm0.03$\nl
NGC 7785&$8.4\pm2.3$&$0.21\pm0.05$&$0.16\pm0.02$&$0.06\pm0.05$&$0.22\pm0.05$\nl
\sidehead{Fornax cluster ellipticals:}
NGC 1336&$15.9\pm3.0$&$-0.32\pm0.04$&$0.13\pm0.04$&$-0.44\pm0.05$&$-0.31\pm0.04$\nl
NGC 1339&$12.7\pm4.8$&$0.12\pm0.07$&$0.22\pm0.03$&$-0.08\pm0.08$&$0.14\pm0.07$\nl
NGC 1351&$17.0\pm3.3$&$-0.10\pm0.05$&$0.16\pm0.03$&$-0.25\pm0.06$&$-0.09\pm0.05$\nl
NGC 1373&$6.3\pm2.0$&$0.13\pm0.08$&$0.13\pm0.03$&$0.01\pm0.08$&$0.14\pm0.08$\nl
NGC 1374&$9.5\pm2.6$&$0.13\pm0.07$&$0.18\pm0.02$&$-0.04\pm0.07$&$0.14\pm0.07$\nl
NGC 1379&$10.9\pm2.9$&$-0.08\pm0.06$&$0.16\pm0.03$&$-0.23\pm0.07$&$-0.07\pm0.06$\nl
NGC 1399&$11.5\pm2.4$&$0.29\pm0.06$&$0.25\pm0.03$&$0.06\pm0.07$&$0.31\pm0.06$\nl
NGC 1404&$9.0\pm2.5$&$0.25\pm0.05$&$0.14\pm0.03$&$0.12\pm0.06$&$0.26\pm0.05$\nl
NGC 1419&$13.7\pm3.2$&$-0.09\pm0.06$&$0.09\pm0.03$&$-0.17\pm0.07$&$-0.08\pm0.06$\nl
NGC 1427&$12.2\pm1.6$&$-0.07\pm0.03$&$0.11\pm0.02$&$-0.17\pm0.04$&$-0.06\pm0.03$\nl
IC 2006&$16.9\pm4.2$&$ 0.06\pm0.06$&$0.16\pm0.03$&$-0.09\pm0.07$&$ 0.07\pm0.06$\nl
\enddata
\end{deluxetable}

Table~\ref{tbl:re8tza} presents derived SSP parameters $(t,\z,\enh)$
and their uncertainties through the \reo{8}\ aperture under the
preferred enrichment model 4 of Paper I.  The quantity \enh\ is
similar to the quantity $[\alpha/{\rm Fe}]$ used by other authors, but
we have fine-tuned the elements in the ``E'' group based on current
knowledge.  The E group in model 4 contains Ne, Na, Mg, Si, S, as well
as C and O; the abundance of these elements is slightly enhanced
relative to the mean.  A ``depressed group'' contains the Fe--peak
elements, while all other elements are held constant (at fixed \z).
See Paper I for further details on element grouping and notation.

The above grouping of elements is based partly on observed elliptical
line-strengths and partly on current nucleosynthetic theory.  The
observed strength of Mg (and Na) in ellipticals strongly implies the
enhancement of O and other $\alpha$-elements, as these elements are
nucleosynthetically linked (\cite{WW95}).  (Note that the nominal
$\alpha$-element Ca seems to belong with the Fe--peak elements in
ellipticals based on its line strengths [\cite{Worthey98};
\cite{TWFBG98}]; this anomaly is unexplained.)  The element C is also
clearly strong in giant ellipticals and is placed in the E group for
that reason (\cite{Worthey98}; Paper I).  On the other hand, the weak
Fe lines of ellipticals suggest a reduction in Fe--peak elements
(\cite{Worthey98}).  All remaining elements have been left unchanged
for lack of information, although in retrospect N should probably have
been grouped in the E group, but this makes little difference to the
final results (see Paper I).

Paper I argued that it is actually incorrect to think of the E
elements as being enhanced in elliptical galaxies; since they dominate
\z\ by mass, their abundance essentially \emph{is} \z.  If \enh\ is
$>0$, it must rather be that the Fe--peak elements are
\emph{depressed} (relative to the average element).  The Fe--peak
elements contribute so little to the overall metallicity (only 8\% at
solar abundance) that changing their abundance by large amounts does
not significantly affect either \eh\ or \z.  Thus, in what follows we
think consistently of the relative depression of the Fe--peak elements
rather than the relative enhancement of the E elements.  Specifically,
if $\enh \ne 0$, then [E/Z] is very slightly positive while [Fe/Z] is
nearly equal to $-\enh$.

Table~\ref{tbl:re8tza} also presents the further quantities \feh\ and
\eh.  These are computed using the equations
\begin{equation}
\feh = \z - A \enh\label{eq:feh}
\end{equation}
and
\begin{equation}
\eh = \z + (1 - A) \enh,
\end{equation}
where $A=0.929$ for enrichment model 4 (see Paper I for details).

\subsection{Global parameters}

\begin{deluxetable}{lrrrrrrcccl}
\tablefontsize{\footnotesize}
\tablenum{2a}
\tablecaption{Distance-independent Quantities\label{tbl:diq}}
\tablewidth{0pt}
\tablehead{&\colhead{$\sigma$}&&&\colhead{$r_e$}&&
\colhead{$a_4/a$}&&&\colhead{Nuclear}&
\\
\colhead{Name}&\colhead{(\kms)}&\colhead{$B_T^O$}&\colhead{$\epsilon$}&
\colhead{$(^{\prime\prime})$}&\colhead{$\langle\mu_e\rangle$}&
\colhead{$\times100$}&\colhead{\vos}&
\colhead{$\Sigma_{\rm SS}$}&\colhead{profile}&\colhead{AGN?}}
\tablecolumns{11}
\startdata
NGC 221&$72\pm\phn3$& 8.72&0.23&39&18.70&0.00&0.89&\nodata&$\setminus$&no\nl
NGC 224&$156\pm\phn4$& 5.58&0.18&\nodata&\nodata&\nodata&0.78&\nodata&$\cap$&no\nl
NGC 315&$321\pm\phn4$&11.87&0.27&55&22.26&-0.30&0.09&\nodata&\nodata&LINER\nl
NGC 507&$262\pm\phn6$&12.13&0.12&77&23.06&\nodata&0.09&\nodata&\nodata&no\nl
NGC 547&$236\pm\phn4$&12.92&0.16&25&22.02&0.00&0.24&\nodata&\nodata&\nodata\nl
NGC 584&$193\pm\phn3$&11.21&0.30&30&20.58&1.50&1.55&2.78&\nodata&\nodata\nl
NGC 636&$160\pm\phn3$&12.22&0.13&19&20.72&0.80&1.04&1.48&\nodata&\nodata\nl
NGC 720&$239\pm\phn5$&11.13&0.39&40&21.16&0.35&0.32&\nodata&$\cap$&\nodata\nl
NGC 821&$189\pm\phn3$&11.72&0.32&36&21.49&2.50&0.70&\nodata&\nodata&no\nl
NGC 1336&$96\pm\phn5$&13.08&0.26&27&22.16&\nodata&\nodata&\nodata&\nodata&\nodata\nl
NGC 1339&$158\pm\phn9$&12.50&0.29&17&20.64&\nodata&1.22&\nodata&\nodata&\nodata\nl
NGC 1351&$157\pm\phn9$&12.48&0.34&26&21.33&\nodata&0.80&\nodata&\nodata&\nodata\nl
NGC 1373&$75\pm\phn4$&14.08&0.23&10&21.00&\nodata&\nodata&\nodata&\nodata&\nodata\nl
NGC 1374&$185\pm10$&12.01&0.09&30&21.26&\nodata&\nodata&\nodata&\nodata&\nodata\nl
NGC 1379&$130\pm\phn7$&11.87&0.03&42&21.79&\nodata&\nodata&\nodata&\nodata&\nodata\nl
NGC 1399&$375\pm21$&10.44&0.10&42&20.68&0.10&0.25&\nodata&$\cap$&\nodata\nl
NGC 1404&$260\pm14$&10.98&0.11&27&20.02&\nodata&\nodata&\nodata&\nodata&\nodata\nl
NGC 1419&$117\pm\phn6$&13.46&0.00&11&20.59&\nodata&\nodata&\nodata&\nodata&\nodata\nl
NGC 1427&$175\pm10$&11.81&0.31&33&21.34&\nodata&0.39&\nodata&\nodata&\nodata\nl
IC 2006&$136\pm\phn8$&12.25&0.10&29&21.45&\nodata&\nodata&\nodata&\nodata&\nodata\nl
NGC 1453&$286\pm\phn4$&12.26&0.17&28&21.47&-0.50&0.62&1.48&\nodata&\nodata\nl
NGC 1600&$315\pm\phn4$&11.83&0.33&47&22.15&-0.75&0.03&\nodata&$\cap$&\nodata\nl
NGC 1700&$227\pm\phn3$&12.01&0.27&24&20.82&0.70&0.59&3.70&$\setminus$&\nodata\nl
NGC 2300&$252\pm\phn3$&11.77&0.16&34&21.42&0.60&0.08&2.85&\nodata&no\nl
NGC 2778&$154\pm\phn3$&13.21&0.21&19&21.60&-0.20&0.74&\nodata&\nodata&\nodata\nl
NGC 3377&$108\pm\phn3$&11.07&0.50&34&20.78&1.05&0.86&1.48&$\setminus$&no\nl
NGC 3379&$203\pm\phn3$&10.18&0.09&35&20.15&0.10&0.72&0.00&$\cap$&LINER?\nl
NGC 3608&$178\pm\phn3$&11.69&0.19&35&21.40&-0.20&0.27&0.00&$\cap$&LINER:\nl
NGC 3818&$173\pm\phn4$&12.47&0.39&21&21.17&2.30&0.93&1.30&\nodata&\nodata\nl
NGC 4261&$288\pm\phn3$&11.36&0.21&39&21.26&-1.30&0.10&1.00&\nodata&LINER\nl
NGC 4374&$282\pm\phn3$&10.01&0.14&52&20.73&-0.40&0.09&2.30&\nodata&LINER\nl
NGC 4472&$279\pm\phn4$& 9.33&0.16&104&21.40&-0.25&0.43&\nodata&$\cap$&Sy2?\nl
NGC 4478&$128\pm\phn2$&12.21&0.19&14&19.87&-0.75&0.84&\nodata&$\setminus$&no\nl
NGC 4489&$47\pm\phn4$&12.88&0.12&32&22.23&-0.20&1.49&\nodata&\nodata&\nodata\nl
NGC 4552&$252\pm\phn3$&10.57&0.07&30&20.22&0.01&0.28&\nodata&$\cap$&trans\nl
NGC 4649&$310\pm\phn3$& 9.70&0.17&74&21.11&-0.35&0.42&\nodata&$\cap$&no\nl
NGC 4697&$162\pm\phn4$&10.07&0.41&75&21.40&1.30&0.71&0.00&$\setminus$&\nodata\nl
NGC 5638&$154\pm\phn3$&12.06&0.08&34&21.58&0.20&0.73&\nodata&\nodata&no\nl
NGC 5812&$200\pm\phn3$&11.83&0.05&22&20.65&0.00&0.52&\nodata&\nodata&\nodata\nl
NGC 5813&$205\pm\phn3$&11.42&0.16&49&21.83&0.01&0.51&\nodata&$\cap$&LINER:\nl
NGC 5831&$160\pm\phn3$&12.31&0.17&27&21.44&0.50&0.19&3.60&\nodata&no\nl
NGC 5846&$224\pm\phn4$&10.91&0.07&83&22.26&0.00&0.10&0.30&\nodata&trans:\nl
NGC 6127&$239\pm\phn4$&12.92&0.06&22&21.60&\nodata&0.11&\nodata&\nodata&\nodata\nl
NGC 6702&$174\pm\phn3$&13.04&0.23&29&22.16&-0.40&0.18&\nodata&\nodata&LINER?\nl
NGC 6703&$183\pm\phn3$&11.97&0.02&24&20.88&0.00&0.30&\nodata&\nodata&LINER?\nl
NGC 7052&$274\pm\phn4$&12.69&0.45&32&22.30&0.01&0.34&\nodata&\nodata&\nodata\nl
NGC 7454&$106\pm\phn3$&12.63&0.35&26&21.60&0.00&0.13&\nodata&\nodata&\nodata\nl
NGC 7562&$248\pm\phn3$&12.37&0.29&25&21.28&0.01&0.06&\nodata&\nodata&\nodata\nl
NGC 7619&$300\pm\phn3$&11.93&0.24&32&21.52&0.30&0.53&0.00&\nodata&no\nl
NGC 7626&$253\pm\phn3$&12.06&0.13&38&21.88&0.01&0.12&2.60&\nodata&LINER?\nl
NGC 7785&$240\pm\phn3$&12.41&0.42&27&21.46&-1.20&0.47&\nodata&\nodata&\nodata\nl
\enddata
\tablecomments{
Col.~(1): Galaxy name. 
Col.~(2): Velocity dispersion within $r_e/8$ aperture from
Gonz\'alez (1993) or central velocity dispersion from Kuntschner (1998).
Col.~(3): Total spheroid $B$ magnitude corrected for Galactic absorption and
redshift; see Table~\ref{tbl:colors} for details.
Col.~(4): Mean ellipticity from $\sim7\arcsec$ to $r_e$, from
Gonz{\'a}lez (1993) or Caon et al.~(1994).
Col.~(5): Effective radius in arc seconds in the Seven Samurai
$\sqrt{ab}$ system from Faber et al.~(1989), from Gonz{\'a}lez (1993),
or from Caon et al.~(1994).
Col.~(6): Mean effective surface brightness inside $r_e$ in $B$
magnitudes per square arc second, from Faber et al.\ (1989) or computed
from values in Caon et al.~(1994).
Col.~(7): Isophotal shape parameter, $a_4/a \times 100$, from Faber et
al.\ (1997), Bender, Burstein \& Faber (1992), and Bender (priv.\
comm.).
Col.~(8): Rotation parameter $\vos=\langle v/\sigma_0\rangle/\langle
v/\sigma_0\rangle_{\rm oblate}$, where $\langle v/\sigma_0\rangle_{\rm
oblate}=[\epsilon/(1-\epsilon)]^{1/2}$, as defined in Bender (1988).
Taken from Faber et al.\ (1997) and Bender, Burstein \& Faber (1992),
or derived from data in Gonz{\'a}lez (1993) and Kuntschner (1998) when
necessary.  The rotational velocity for NGC 4489 is taken from
Prugniel \& Simien (1996), and its \vos\ should be considered an upper
limit.  NGC 1427 has a kinematically decoupled core (e.g., Kuntschner
1998); its value is an upper limit and may be much closer to zero.
Col.~(9): Morphological disturbance parameter from Schweizer \&
Seitzer (1992).
Col.~(10): Nuclear profile shape from Faber et al.\ (1997):
``$\cap$'' denotes core; ``$\setminus$'' denotes power-law.
Col.~(11): AGN detection and classification from Ho,
Filippenko \& Sargent (1997):  Sy=Seyfert; trans=intermediate AGN
(LINER/\ion{H}{2} nucleus); LINER=LINER; no=no AGN detected.  ``:''
denotes uncertain classification, ``?'' denotes highly uncertain
classification.}
\end{deluxetable}

\begin{deluxetable}{lrccccrcc}
\tablefontsize{\footnotesize}
\tablenum{2b} 
\tablecaption{Distance-dependent Quantities\label{tbl:ddq}}
\tablewidth{0pt}
\tablehead{&\colhead{$cz$}&&&\colhead{$\log r_e$}&
\colhead{$\log I_e$}&\colhead{$\log
M$}&\colhead{$M/L_B$} \\
\colhead{Name}&\colhead{(\kms)}&\colhead{$(m-M)_{CMB}$}&
\colhead{$M_B$}&\colhead{(pc)}&\colhead{($L_{\odot}\;\rm pc^{-2}$)}&
\colhead{($M_{\odot}$)}&\colhead{($M_{\odot}/L_{\odot}$)}}
\tablecolumns{8}
\startdata
NGC 221&$-204\pm\phn7$&24.63&$-15.91$&2.20&3.32&\phn8.58&\phn2.27\nl
NGC 224&$-300\pm\phn7$&24.48&\nodata&\nodata&\nodata&\nodata&\nodata\nl
NGC 315&$4942\pm\phn6$&34.02&$-22.15$&4.23&1.90&11.91&11.26\nl
NGC 507&$4908\pm11$&33.87&$-21.74$&4.35&1.58&11.85&11.99\nl
NGC 547&$5468\pm\phn6$&34.12&$-21.20$&3.91&1.99&11.32&10.25\nl
NGC 584&$1866\pm\phn6$&31.60&$-20.39$&3.48&2.57&10.72&\phn4.84\nl
NGC 636&$1860\pm\phn6$&32.45&$-20.23$&3.45&2.51&10.53&\phn4.04\nl
NGC 720&$1741\pm11$&32.29&$-21.16$&3.75&2.34&11.17&\phn6.91\nl
NGC 821&$1730\pm\phn7$&31.99&$-20.27$&3.64&2.20&10.86&\phn7.47\nl
NGC 1336&$1439\pm11$&31.52&$-18.44$&3.42&1.94&10.05&\phn5.92\nl
NGC 1339&$1355\pm12$&31.52&$-19.02$&3.22&2.54&10.28&\phn6.28\nl
NGC 1351&$1529\pm13$&31.52&$-19.04$&3.40&2.27&10.46&\phn7.65\nl
NGC 1373&$1341\pm10$&31.52&$-17.44$&2.99&2.40&\phn9.41&\phn3.35\nl
NGC 1374&$1349\pm13$&31.52&$-19.51$&3.47&2.30&10.67&\phn8.63\nl
NGC 1379&$1360\pm11$&31.52&$-19.65$&3.61&2.08&10.51&\phn4.96\nl
NGC 1399&$1431\pm28$&31.52&$-21.08$&3.61&2.53&11.43&14.85\nl
NGC 1404&$1923\pm17$&31.52&$-20.54$&3.42&2.79&10.92&\phn6.04\nl
NGC 1419&$1574\pm10$&31.52&$-18.06$&3.03&2.56&\phn9.83&\phn5.08\nl
NGC 1427&$1416\pm10$&31.52&$-19.71$&3.51&2.26&10.66&\phn7.56\nl
IC 2006&$1371\pm12$&31.52&$-19.27$&3.45&2.22&10.39&\phn5.75\nl
NGC 1453&$3886\pm\phn6$&33.59&$-21.33$&3.85&2.21&11.43&10.34\nl
NGC 1600&$4688\pm\phn8$&34.06&$-22.23$&4.17&1.94&11.83&11.25\nl
NGC 1700&$3895\pm\phn7$&33.31&$-21.30$&3.73&2.47&11.11&\phn4.75\nl
NGC 2300&$1938\pm\phn7$&32.15&$-20.38$&3.65&2.23&11.12&12.25\nl
NGC 2778&$2060\pm\phn7$&31.88&$-18.67$&3.34&2.16&10.38&10.94\nl
NGC 3377&$ 724\pm\phn7$&30.33&$-19.26$&3.28&2.49&10.02&\phn2.89\nl
NGC 3379&$ 945\pm\phn7$&30.20&$-20.02$&3.27&2.74&10.55&\phn5.88\nl
NGC 3608&$1222\pm\phn7$&31.88&$-20.19$&3.61&2.24&10.77&\phn6.60\nl
NGC 3818&$1708\pm10$&32.88&$-20.41$&3.58&2.33&10.73&\phn5.30\nl
NGC 4261&$2238\pm\phn7$&32.58&$-21.22$&3.79&2.30&11.38&\phn9.87\nl
NGC 4374&$1060\pm\phn6$&31.40&$-21.39$&3.68&2.51&11.25&\phn7.50\nl
NGC 4472&$ 980\pm10$&31.14&$-21.81$&3.93&2.24&11.49&\phn7.67\nl
NGC 4478&$1365\pm\phn7$&31.37&$-19.16$&3.11&2.85&\phn9.99&\phn2.64\nl
NGC 4489&$ 970\pm10$&31.34&$-18.46$&3.46&1.91&\phn9.47&\phn1.39\nl
NGC 4552&$ 364\pm\phn7$&31.01&$-20.44$&3.36&2.71&10.83&\phn7.77\nl
NGC 4649&$1117\pm\phn6$&31.21&$-21.51$&3.80&2.36&11.45&\phn9.87\nl
NGC 4697&$1307\pm10$&30.43&$-20.36$&3.65&2.24&10.73&\phn4.97\nl
NGC 5638&$1649\pm\phn6$&32.18&$-20.12$&3.65&2.17&10.70&\phn5.23\nl
NGC 5812&$1929\pm\phn7$&32.23&$-20.40$&3.47&2.54&10.74&\phn5.65\nl
NGC 5813&$1954\pm\phn7$&32.62&$-21.20$&3.90&2.07&11.19&\phn6.61\nl
NGC 5831&$1655\pm\phn5$&32.25&$-19.94$&3.57&2.22&10.64&\phn6.05\nl
NGC 5846&$1714\pm\phn5$&32.06&$-21.15$&4.02&1.90&11.38&\phn8.96\nl
NGC 6127&$4700\pm10$&33.95&$-21.03$&3.82&2.16&11.24&\phn8.77\nl
NGC 6702&$4728\pm\phn5$&33.59&$-20.55$&3.87&1.94&11.01&\phn6.97\nl
NGC 6703&$2403\pm\phn7$&32.18&$-20.21$&3.50&2.45&10.69&\phn5.49\nl
NGC 7052&$4672\pm\phn8$&33.83&$-21.14$&3.96&1.88&11.50&15.96\nl
NGC 7454&$2051\pm\phn7$&31.97&$-19.34$&3.49&2.16&10.21&\phn3.63\nl
NGC 7562&$3608\pm\phn5$&33.87&$-21.50$&3.86&2.29&11.31&\phn6.42\nl
NGC 7619&$3762\pm\phn5$&33.70&$-21.77$&3.93&2.19&11.55&\phn9.91\nl
NGC 7626&$3405\pm\phn4$&33.09&$-21.03$&3.88&2.05&11.36&10.95\nl
NGC 7785&$3808\pm\phn5$&33.32&$-20.91$&3.78&2.22&11.21&\phn8.47\nl
\enddata
\tablecomments{Col.~(1): Galaxy name.  Col.~(2): Heliocentric radial
velocity from Gonz{\'a}lez (1993) or Kuntschner (1998).  Col.~(3):
CMB-frame distance modulus from SBF measurements (Tonry et al., in
prep.) or flow-corrected models (Tonry et al., priv.~comm.).
Col.~(4): Absolute $B$ magnitude, computed from $B_T^O$ in
Table~\ref{tbl:diq}, col.~(4) and the distance in col.~(3) here.
Col.~(5): Logarithm of the effective radius in parsecs.  Col.~(6):
Logarithm of mean $B$ surface brightness inside $r_e$ in solar
luminosities per parsec$^2$ ($I_e = 10^{-0.4(\langle\mu_e\rangle -
27.0)}$; see Bender, Burstein \& Faber 1992).  Col.~(7): Logarithm of
the galaxy mass within the effective radius, in solar masses.
Computed as $M=465\sigma_0^2 r_e\ M_{\odot}$ (Burstein et al.\ 1997).
Col.~(8): Mass-to-light ratio within the effective radius in the $B$
band.  Computed as $M/L = 146 \sigma_0^2/(I_e r_e)\
M_{\odot}/L_{\odot}$ (Gonz{\'a}lez 1993; Burstein et al.\ 1997).}
\end{deluxetable}

Structural parameters are presented in Tables~\ref{tbl:diq} and
\ref{tbl:ddq}.  Table~\ref{tbl:diq} gives distance-independent
quantities: velocity dispersions (from G93 and K98), $B_T^O$
magnitudes (Section~\ref{sec:colors}), mean ellipticities and
effective radii in arc seconds (collected from the literature and
homogenized by G93), mean effective surface brightnesses, isophotal
shape parameters $a_4/a$, rotation parameters \vos, morphological
disturbance parameters $\Sigma_{\rm SS}$ (\cite{Schweizer90},
\cite{SS92}), nuclear profile shapes (power-law or core;
\cite{Faber97}), and presence and type of AGN activity, if any.
Table~\ref{tbl:ddq} presents distance-dependent quantities: redshifts
(repeated from Table 1 of Paper I), distance moduli from SBF
measurements (Tonry et al., in prep.)  or flow-corrected distances
from Tonry et al.~(priv.~comm.), absolute magnitudes using SBF
distances, effective radii in parsecs, mean effective surface
brightnesses in solar units (not distance-dependent but needed in the
computation of mass-to-light ratios), galaxy masses in solar masses,
and mass-to-(blue)-light ratios in solar units.  Many of these
quantities will be used in future papers.  Details and references are
given in the footnotes to the tables.

\subsection{Magnitudes and colors}\label{sec:colors}

\begin{deluxetable}{lrcccccc}
\setcounter{table}{2}
\tablewidth{0pt}
\tablecaption{$UBV$ Photometry from Literature\label{tbl:colors}}
\tablehead{\colhead{Name}&\colhead{$B_T^O$}&\colhead{$(U-V)_T^O$}
&\colhead{$(U-V)_e^O$}&\colhead{$(U-V)_8^O$}&\colhead{$(B-V)_T^O$}
&\colhead{$(B-V)_e^O$}&\colhead{$(B-V)_8^O$}}
\startdata
NGC 221& 8.72&1.28&1.31&1.46&0.88&0.89&0.94\nl
NGC 224& 5.58&0.99&\nodata&\nodata&0.68&\nodata&\nodata\nl
NGC 315&11.87&1.49&1.53&1.68&0.93&0.96&1.01\nl
NGC 507&12.13&1.41&1.47&1.62&0.91&0.93&0.98\nl
NGC 547&12.92&\nodata&1.43&1.58&\nodata&0.95&1.01\nl
NGC 584&11.21&1.38&1.44&1.59&0.91&0.92&0.97\nl
NGC 636&12.22&1.36&1.41&1.56&0.90&0.91&0.96\nl
NGC 720&11.13&1.44&1.51&1.66&0.96&0.97&1.02\nl
NGC 821&11.72&\nodata&1.52&1.67&0.93&0.94&0.99\nl
NGC 1336&13.08&1.07&1.13&1.28&0.82&0.83&0.88\nl
NGC 1339&12.50&1.41&1.46&1.61&0.92&0.93&0.98\nl
NGC 1351&12.48&1.23&1.33&1.48&0.87&0.91&0.96\nl
NGC 1373&14.08&1.18&\nodata&\nodata&0.85&\nodata&\nodata\nl
NGC 1374&12.01&1.38&1.44&1.59&0.91&0.93&0.98\nl
NGC 1379&11.87&1.26&1.32&1.47&0.88&0.90&0.95\nl
NGC 1399&10.44&1.46&1.54&1.69&0.95&0.97&1.02\nl
NGC 1404&10.98&1.52&1.55&1.70&0.95&0.97&1.02\nl
NGC 1419&13.46&1.21&1.26&1.41&0.88&0.89&0.94\nl
NGC 1427&11.81&1.33&1.35&1.50&0.90&0.91&0.96\nl
IC 2006&12.25&1.31&1.39&1.54&0.91&0.94&0.99\nl
NGC 1453&12.26&1.53&1.58&1.73&0.96&0.98&1.03\nl
NGC 1600&11.83&1.50&1.57&1.72&0.95&0.97&1.02\nl
NGC 1700&12.01&1.40&1.46&1.61&0.91&0.92&0.97\nl
NGC 2300&11.77&1.66&1.68&1.83&1.01&1.02&1.07\nl
NGC 2778&13.21&1.42&1.47&1.62&0.91&0.94&0.99\nl
NGC 3377&11.07&1.14&1.26&1.41&0.84&0.87&0.92\nl
NGC 3379&10.18&1.46&1.52&1.67&0.94&0.96&1.01\nl
NGC 3608&11.69&1.33&1.44&1.59&0.93&0.95&1.00\nl
NGC 3818&12.47&\nodata&1.46&1.61&0.92&0.93&0.98\nl
NGC 4261&11.36&1.50&1.57&1.72&0.97&0.98&1.03\nl
NGC 4374&10.01&1.44&1.49&1.64&0.94&0.95&1.00\nl
NGC 4472& 9.33&1.51&1.57&1.72&0.95&0.97&1.02\nl
NGC 4478&12.21&1.33&1.35&1.50&0.88&0.89&0.94\nl
NGC 4489&12.88&1.10&1.23&1.38&0.83&0.86&0.91\nl
NGC 4552&10.57&1.47&1.55&1.70&0.94&0.96&1.01\nl
NGC 4649& 9.70&\nodata&1.61&1.76&0.95&0.98&1.03\nl
NGC 4697&10.07&1.28&1.37&1.52&0.89&0.92&0.97\nl
NGC 5638&12.06&1.34&1.39&1.54&0.91&0.92&0.97\nl
NGC 5812&11.83&\nodata&1.51&1.66&0.94&0.94&0.99\nl
NGC 5813&11.42&1.46&1.51&1.66&0.94&0.95&1.00\nl
NGC 5831&12.31&1.47&1.49&1.64&0.92&0.93&0.98\nl
NGC 5846&10.91&1.41&1.52&1.67&0.96&0.98&1.03\nl
NGC 6127&12.92&\nodata&\nodata&\nodata&0.96&0.97&1.02\nl
NGC 6702&13.04&1.37&1.49&1.64&0.89&0.94&0.99\nl
NGC 6703&11.97&1.40&1.46&1.61&0.91&0.93&0.98\nl
NGC 7052&12.69&\nodata&\nodata&\nodata&\nodata&\nodata&\nodata\nl
NGC 7454&12.63&1.19&1.29&1.44&0.89&0.90&0.95\nl
NGC 7562&12.37&1.58&1.61&1.76&0.98&0.99&1.04\nl
NGC 7619&11.93&1.51&1.59&1.74&0.96&0.98&1.03\nl
NGC 7626&12.06&1.52&1.56&1.71&0.98&0.99&1.04\nl
NGC 7785&12.41&1.48&1.59&1.74&0.96&0.97&1.02\nl
\enddata
\tablecomments{All colors and magnitudes have been corrected for
Galactic absorption and redshift.  Col.~(1): Galaxy name.  Col.~(2):
Total $B$ magnitude from the RC3 for all galaxies except NGC 547 (Poulain
\& Nieto 1994) and the NGC 224 bulge (Faber et al.~1997).
Col.~(3): Total $(U-V)$ color from RC3.  Col.~(4): Effective $(U-V)$
color at $r_e$ from RC3 for all galaxies except NGC 547, NGC 3818, NGC
5812 (Poulain \& Nieto 1994), and NGC 821 (Poulain 1988).  Col.~(5):
Central $(U-V)$ color within $r_e/8$ extrapolated inward from
$(U-V)_e^O$ using mean logarithmic radial gradients from Peletier
et al.~(1990) and Goudfrooij et al.~(1994) (Section~\ref{sec:colors}).
Col.~(6): Total $(B-V)$ color from RC3.  Col.~(7): Effective $(B-V)$
color at $r_e$ from RC3 for all galaxies except NGC 547 (Poulain \&
Nieto 1994).  Col.~(8): Central $(B-V)$ color within $r_e/8$
extrapolated inward from $(B-V)_e^O$ using the mean logarithmic radial
gradient from Goudfrooij et al.~(1994) (Section~\ref{sec:colors}).}
\end{deluxetable} 

Table~\ref{tbl:colors} presents $B_T^O$, \umv, and \bmv\ in various
apertures for all galaxies except NGC 7052, for which no published
global photometry was found.  These values are corrected for Galactic
absorption and redshift (but not internal extinction) following the
precepts of the RC3.  ``Total'' and ``effective'' colors are drawn
from the RC3, Poulain (1988), or Poulain \& Nieto (1994) as
appropriate.  A ``central'' color through \reo{8}\ is computed by
taking effective colors and correcting them inward using the average
color gradients of early-type galaxies from Peletier et al.\ (1990)
and Goudfrooij et al.\ (1994).  The mean \bmv\ color gradient is taken
from Goudfrooij et al.\ (1994):
\begin{equation}
\frac{\Delta(B-V)}{\Delta(\log r)} = -0.06 \pm 0.01\ \mathrm{mag/dex}
\end{equation}
(using 53 galaxies).  The mean \umb\ gradient is from Peletier et al.\
(1990):
\begin{equation}
\frac{\Delta(U-B)}{\Delta(\log r)} = -0.11 \pm 0.03\ \mathrm{mag/dex},
\end{equation}
for a mean \umv\ gradient of
\begin{equation}
\frac{\Delta(U-V)}{\Delta(\log r)} = -0.17 \pm 0.03\ \mathrm{mag/dex}.
\end{equation}
This is consistent with estimates by Peletier, Valentijn \& Jameson
(1990) and the combined results of Franx, Illingworth \& Heckman
(1989) and Goudfrooij et al.\ (1994).  The \reo{8}\ colors are then
computed as
\begin{equation}
\umv_{\reo{8}}^O=\umv_e^O + 0.15
\end{equation}
and
\begin{equation}
\bmv_{\reo{8}}^O=\bmv_e^O + 0.05.
\end{equation}

\section{The manifold of stellar populations of local elliptical
galaxies}\label{sec:results}

\subsection{Principal component analysis}\label{sec:pca}

This section explores the general landscape of correlations among
central SSP-equivalent population parameters (age, metallicity,
enhancement ratio, and iron abundance) and the corresponding
structural parameters of the parent galaxies.  We show below that,
among the structural variables, only velocity dispersion correlates
significantly with the stellar populations.  Furthermore, \feh\ can be
derived from \z\ and \enh.  Hence, this section explores the space of
the four remaining significant variables $t$, \z, \enh, and $\sigma$.

\begin{deluxetable}{crrrr}
\tablefontsize{\normalsize}
\tablecaption{Principal Component Analysis\label{tbl:pca}}
\tablewidth{0pt}
\tablehead{\colhead{Variable}&\colhead{PC1}&\colhead{PC2}&\colhead{PC3}
&\colhead{PC4}}
\startdata
$\sigma^{\prime}$&0.64&0.16&$-$0.68&0.34\cr
$t^{\prime}$&0.11&0.76&$-$0.03&$-$0.64\cr
$z^{\prime}$&0.42&$-$0.62&$-$0.09&$-$0.66\cr
$e^{\prime}$&0.64&0.11&0.73&0.21\cr
\cr
Eigenvalue&2.02&1.63&0.27&0.07\cr
Percentage of variance&50&41&7&2\cr
Cumulative percentage&50&91&98&100\cr
\enddata
\tablecomments{Primed variables are ``reduced'' versions of the
corresponding variables with zero mean and unit variance:
\begin{eqnarray*}
\sigma^{\prime}&=&(\log\sigma - 2.27)/1.29,\\
t^{\prime}&=&(\logt - 0.88)/1.82,\\
z^{\prime}&=&(\z - 0.21)/1.29,\\
e^{\prime}&=&(\enh - 0.18)/0.47.
\end{eqnarray*}}
\end{deluxetable}

\begin{figure*}
\plotone{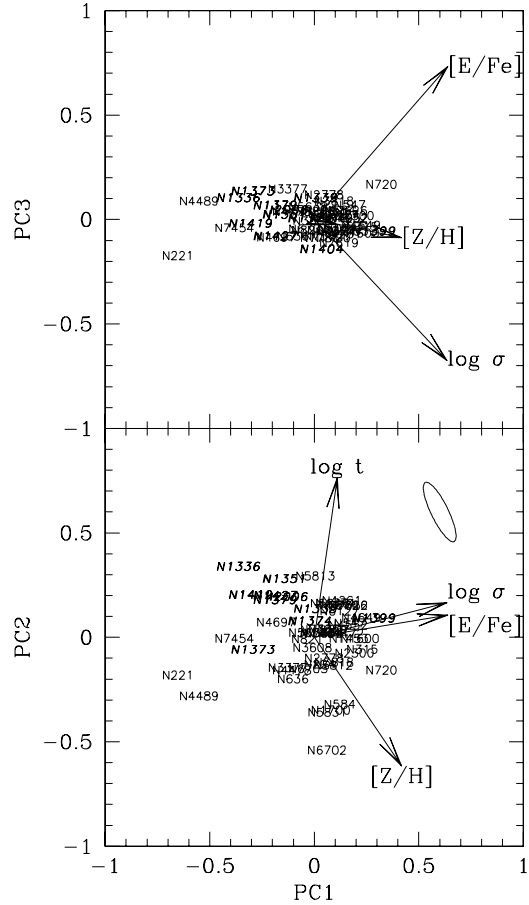}
\caption{The metallicity ``hyperplane'' of stellar populations of
local elliptical galaxies.  Fornax ellipticals are shown in bold,
slanted type; G93 ellipticals are shown in roman type.  Ellipticals
populate a plane in metallicity hyperspace, \tzas.  The lower panel
shows the plane face on.  Projections of the four basic variables are
shown as arrows in the direction of increase (for $\logt$, this arrow
points in the direction of older galaxies).  Velocity dispersion and
enhancement ratio dominate the first principal component, while age
and metallicity dominate the second.  The third and fourth principal
components contribute less than 10\% to the overall variance in \tzas\
space; the ``long axis'' (PC1--PC3) of the hyperplane is shown in the
upper panel.  A $1\sigma$ error ellipse typical of the G93 sample is
shown in the upper right corner of the lower panel.\label{fig:pca}}
\end{figure*}

As an exploratory means of finding the number of independent
parameters in this four-dimensional space, we have performed a
principal component analysis (PCA; see, e.g., \cite{Faber73}) on the
four variables \logt, \z, \enh, and $\log\sigma$.  The results are
presented in Table~\ref{tbl:pca}, where it is shown that the first two
principal components contain 91\% of the variance.  Thus, to high
accuracy, these local ellipticals are confined to a
\emph{two-dimensional surface}, which we propose to call the
``metallicity hyperplane.''  Figure~\ref{fig:pca} shows edge-on and
face-on views of this plane; $\log\sigma$ and \enh\ are the primary
contributors to the first principal component, while $t$ and \z\ drive
the second principal component.

The face-on view of the plane is instructive.  First, \enh\ and
$\sigma$ are nearly coincident.  This is equivalent to saying that one
can substitute for the other, i.e., that they are highly correlated.
Second, $t$, $\sigma$, and \z\ are all moderately orthogonal to one
another, and therefore any one of them can be reasonably well
represented by a linear combination of the other two.  We choose to
regard $\sigma$ and $t$ as independent (see below) and to express \z\
and \enh\ in terms of them.  Hence, to the extent that the thickness
of the plane can be ignored, we predict the following linear
relations: $\z = f(\logt,\log\sigma)$ and $\enh = g(\log\sigma)$.
These are confirmed below.  In summary, to present accuracy and based
on \hbeta, \mgb, and \fe\ alone, the stellar populations of these
local ellipticals are basically a \emph{two-parameter family}
determined mainly by velocity dispersion, $\sigma$, and SSP-equivalent
age, $t$.

The choice of $\sigma$ and $t$ as independent variables is not
mandated by principal components, which only reveals correlations but
cannot show which parameters are fundamental.  The dispersion $\sigma$
was chosen as one independent parameter because it is external to the
stellar populations and might plausibly play a causal role in their
formation.  The selection of $t$ as the second parameter is less
obvious.  However, since \z\ and \enh\ evolve as stars form, it seems
natural to specify them as functions of time rather than the other way
round.  In the end, the choice of $\sigma$ and $t$ as the physically
meaningful, ``independent'' variables is somewhat arbitrary.

\subsection{The \zp-plane}\label{sec:results_tzs}

\begin{figure*}
\plotone{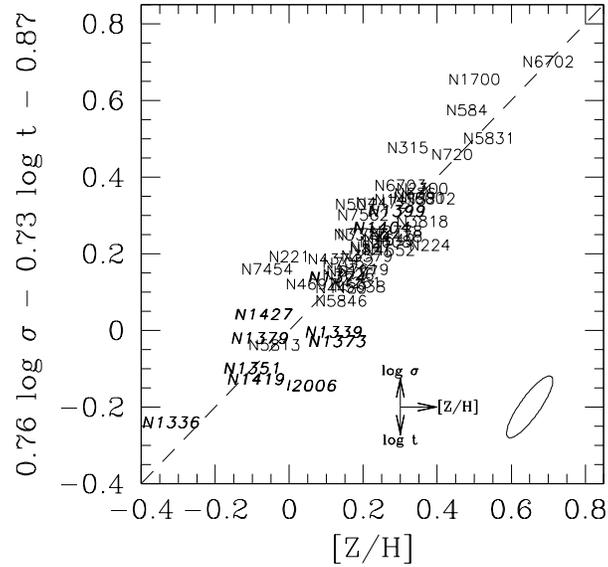}
\caption{An edge-on view of the \zp-plane in hyperspace (points as in
Figure~\ref{fig:re8m}).  The dashed line is the line defining the
plane (Eq.~\ref{eq:tzs}).  Vectors of $\Delta\log\sigma=+0.1$,
$\Delta\logt=+0.1$ (i.e., 26\% older), and $\Delta\z=+0.1$ dex are
shown at bottom, along with a typical error ellipse for the G93
sample.
\label{fig:tzs_edge}}
\end{figure*}

\begin{figure*}
\plotonet{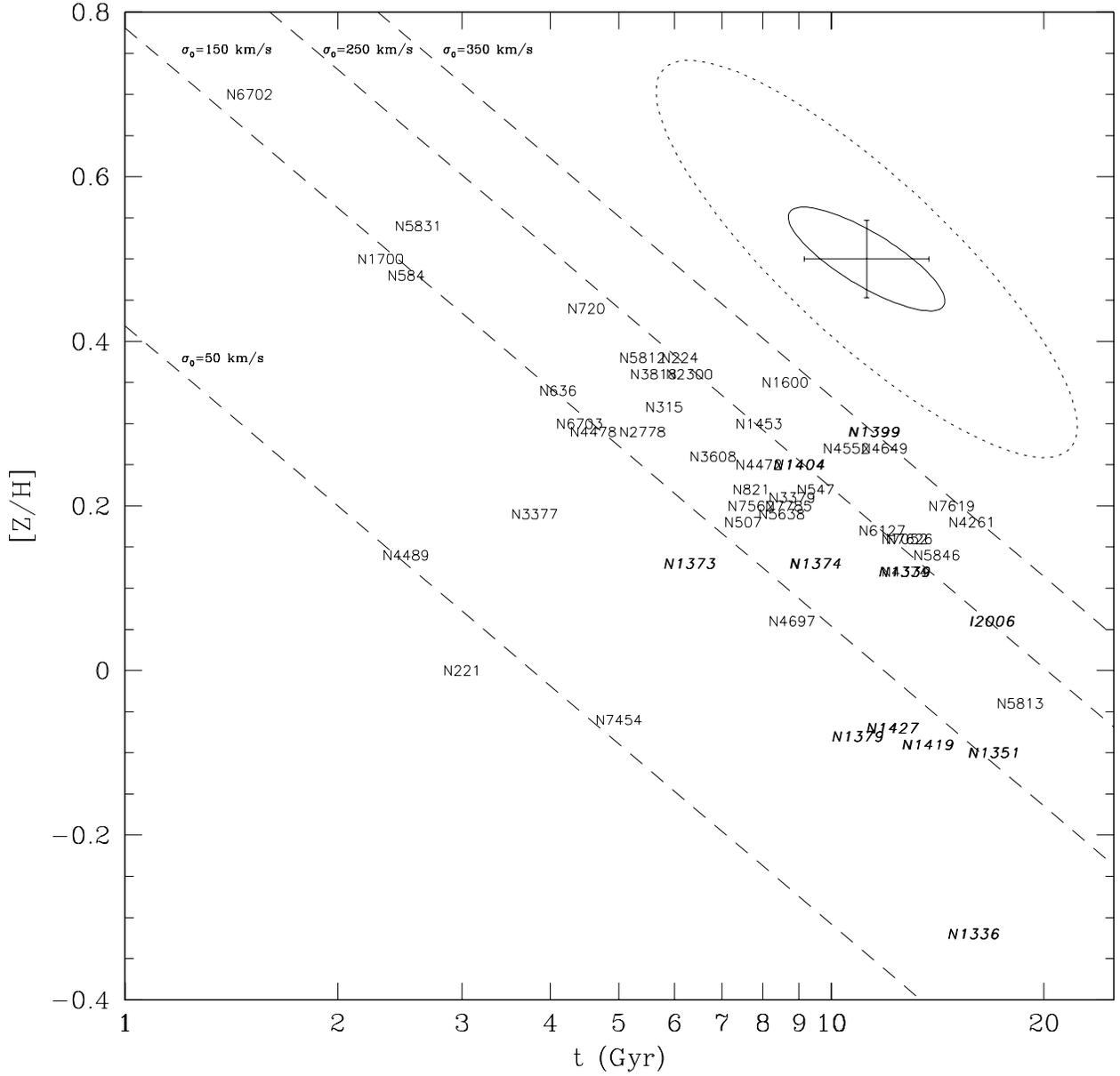}
\caption{A face-on view of the \zp-plane in hyperspace (points as in
Figure~\ref{fig:re8m}).  At fixed velocity dispersion (dashed lines),
younger galaxies have higher metallicities than older galaxies.  The
solid error ellipse in the upper right-hand corner is typical of the
G93 sample; the dotted ellipse is typical of the highest quality data
in the Lick/IDS galaxy sample (\cite{TWFBG98}).  The slope of the
error ellipses is nearly identical to the of lines of constant
velocity dispersion, indicating that poor data can masquerade as real
trends.\label{fig:tz}}
\end{figure*}

Fitting directly now for the planar function $\z\ = f(\logt,\log\sigma)$,
we find:
\begin{eqnarray}
\z & = & \phm{\pm} 0.76\,\log\sigma  - 0.73 \,\logt - 0.87,\label{eq:tzs}\\
   &   & \pm 0.13\phm{\,\log\sigma}\pm 0.06\phm{\,\logt}\pm 0.30\nonumber
\end{eqnarray}
with an RMS residual of 0.09 dex in \z.  (The coefficients have been
determined using the ``orthogonal fit'' procedure of J{\o}ergensen,
Franx \& Kjaergaard 1996, as coded by D. Kelson; the errors have been
estimated using a bootstrap of 1000 replacement samples.)  A similar
plane was found previously by Trager (1997) for the G93 sample using
an older version of SSP parameters that solved for \enh\ rather
crudely; essentially the same results were obtained.  An edge-on view
of this plane is shown in Figure~\ref{fig:tzs_edge}, and the face-on
view is shown in Figure~\ref{fig:tz}.  We call this plane the
``\zp-plane.''

We stated above that $\sigma$ is the only structural variable that
correlates with stellar population parameters.  More precisely, we
mean that adding more structural parameters to fits of the form $\z =
f(\logt,\log\sigma,\log r_e,\log I_e)$ (where $r_e$ is effective
radius and $I_e$ is effective surface brightness) does not
significantly reduce the scatter in \z.  While \z\ should correlate
with mass or luminosity through its correlation with $\sigma$,
substituting mass or luminosity for ($\log\sigma$,$\log r_e$) and
$(\log\sigma,\log r_e,\log I_e)$ respectively in the fits actually
increases the scatter in \z.  This implies that the basic correlation
is through $\sigma$.

The existence of the \zp-plane says that there exists an
\emph{age--metallicity relation} for each value of $\sigma$.  Contours
of constant $\sigma$ are shown in Figure~\ref{fig:tz} and have slope
$\Delta\logt = -1.4\Delta\z$.  This is very close to the ``3/2
relation'' of Worthey (1992, 1994), which expresses trajectories in
$\logt$--\z\ space along which colors and line strengths remain
roughly constant.  Thus, following Trager (1997), we predict that line
strengths should be constant along trajectories of constant $\sigma$
in the \zp-plane, an important conclusion to which we will return in
Section~\ref{sec:mgsigma}.

\subsection{The \enh--$\sigma$ relation}\label{sec:results_sa}

\begin{figure*}
\plotone{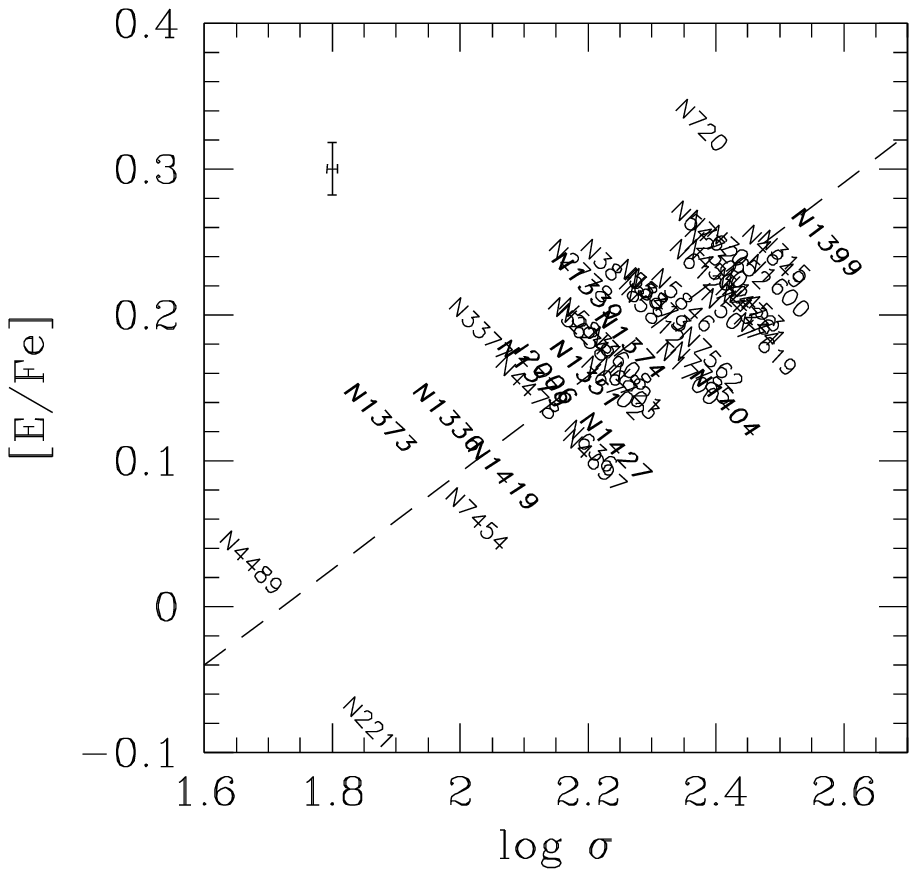}
\caption{The \enh--$\sigma$ relation for local ellipticals (points as
in Figure~\ref{fig:re8m}).  High values of \enh\ actually reflect low
values of [Fe/Z] rather than high [E/Z] (see text).  The dashed line
is a least-squares fit to the relation of the form
$\enh=0.33\log\sigma-0.58$.\label{fig:sa}}
\end{figure*}

PCA analysis indicates that the enhancement ratio, \enh, is closely
coupled to the velocity dispersion, with \enh\ increasing as $\sigma$.
Figure~\ref{fig:sa} confirms this close relationship.  The dashed line
is a linear least-squares fit of the form
\begin{eqnarray}
\enh &=&\phm{\pm}0.33\,\log\sigma - 0.58,\label{eq:sa}\\
     & &\pm0.01\phm{\,\log\sigma}\pm0.01\nonumber
\end{eqnarray}
with an RMS residual of 0.05 dex.  Adding other structural parameters
($\log r_e$ and $\log I_e$) to the fit again does not reduce the
scatter significantly, nor does replacing ($\log\sigma$,$\log r_e$)
with mass. Replacing ($\log\sigma$,$\log r_e$,$\log I_e$) with
luminosity---i.e., fits of the form $\enh = f(\log L)$---actually
increases the scatter slightly.  Thus, although \enh\ obviously
correlates loosely with other structural variables such as mass and
luminosity, the basic correlation is through $\sigma$.  It will be
noted that outlying galaxies from the \enh--$\sigma$ relation also lie
off the plane in the upper panel of Figure~\ref{fig:pca}.  Hence, from
Table~\ref{tbl:pca}, scatter in the \enh--$\sigma$ relation must
reflect the role of PC3 in thickening the hyperplane.  The scatter is
larger than the error bar in Figure~\ref{fig:sa}, indicating that
\enh\ does not correlate perfectly with $\sigma$; the same point was
made also by Kuntschner (1998).  Clearly, the hyperlane has some
finite thickness, and the statement that the galaxies are a
two-dimensional manifold is only approximate.

\subsection{The Fe-plane}\label{sec:results_tf}

\begin{figure*}
\plotone{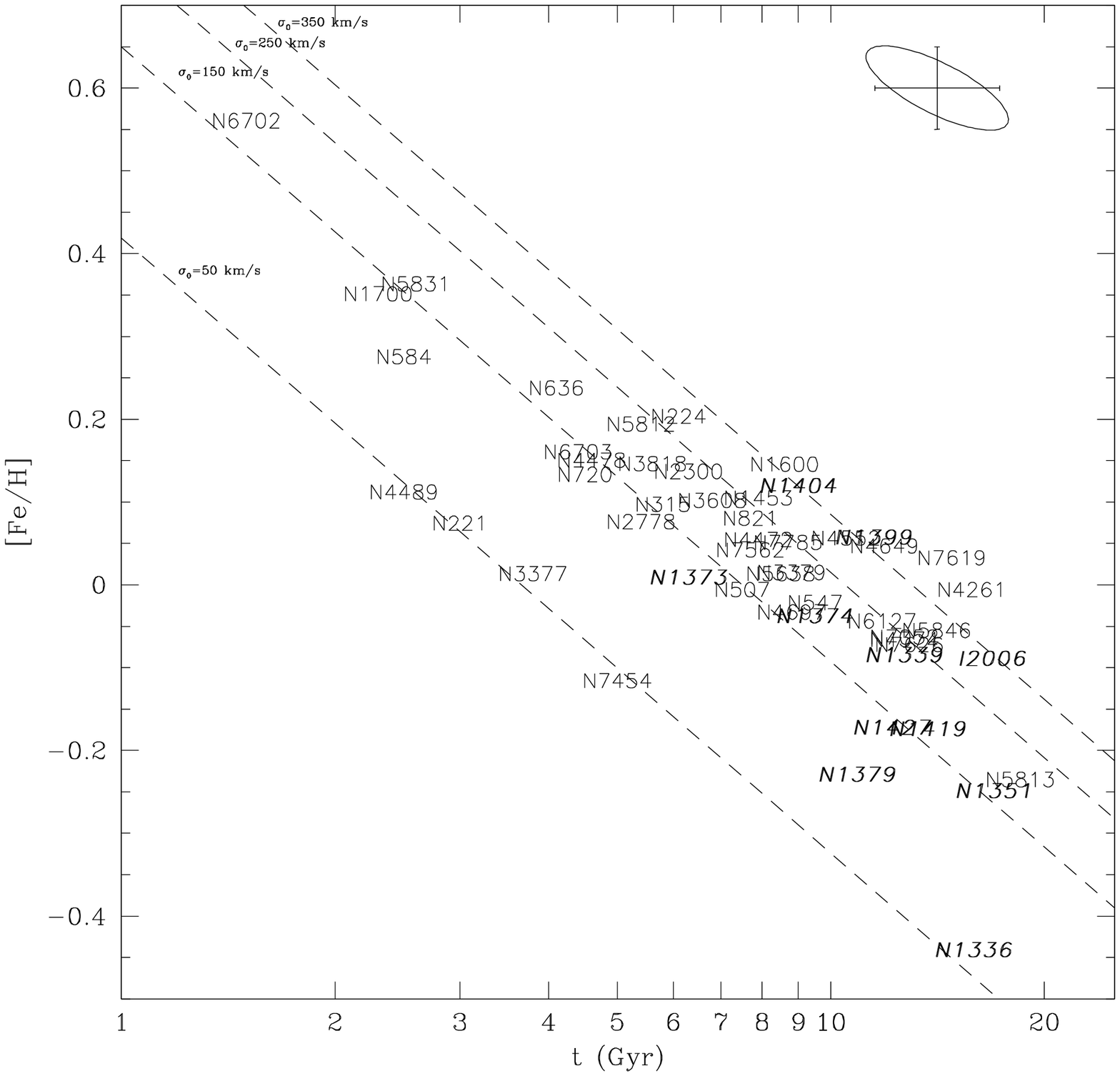}
\caption{A face-on view of the Fe-plane (points as in
Figure~\ref{fig:re8m}).  Younger ellipticals have higher \feh\ than
older ellipticals.  The dashed lines are loci of constant velocity
dispersion.  A typical set of error bars for the G93 sample is shown
in the upper right-hand corner; correlated errors are in the same
direction as the error ellipse in Figure~\ref{fig:tz}.\label{fig:tf}}
\end{figure*}

For completeness we also plot \feh\ as a function of $t$ and $\sigma$
in Figure~\ref{fig:tf}.  Since \feh\ is closely equal to $\z - \enh$
and \enh\ is a function of $\sigma$ only, we predict a plane analogous
to the \zp-plane, but with different slope.  Indeed, such a plane is
found, with equation:
\begin{eqnarray}
\feh & = & \phm{\pm} 0.48\,\log\sigma  - 0.74 \,\logt - 0.40,\label{eq:tfs}\\
     &   & \pm 0.12\phm{\,\log\sigma}\pm 0.09\phm{\,\logt}\pm 0.25\nonumber
\end{eqnarray}
and with an RMS residual of 0.08 dex.  \feh\ is even tighter vs. age
than \z\ (compare Figure~\ref{fig:tf} with Figure~\ref{fig:tz}).  This
tightness is due to the dependence of \enh\ on $\sigma$, which causes
Fe to rise more slowly than \z\ vs. $\sigma$, and thus compresses the
spread in Fe at fixed time.  Mathematically, the Fe-plane is
``flatter'' in velocity dispersion than the \zp-plane.

\subsection{The effect of observational errors}~\label{sec:errors}

It is important to examine the role that observational errors play in
creating the above correlations, particularly the \zp- and Fe-planes.
From Figure~\ref{fig:re8m}, it is evident that an error in any one of
the observed quantities \mgb, \fe, or \hbeta\ will cause correlated
errors in the output quantities \z, \enh, and $t$.  However, \hbeta\
is the most critical index, and errors in it are the most dangerous.
Moving \hbeta\ up in Figure~\ref{fig:re8m} causes age to decline and
\z\ to increase (\enh\ is less affected).  This correlated error is
responsible for the long axis of the tilted error ellipses in the two
plane diagrams, Figures~\ref{fig:tz} and ~\ref{fig:tf}.  Note that
these ellipses point almost directly parallel to the claimed trends in
age at fixed $\sigma$.  Note further that the error ellipse in
Figure~\ref{fig:tzs_edge} is parallel to the edge-on view of the
\zp-plane, indicating that errors do not significantly broaden the
plane (the same is true of the Fe-plane though no edge-on view is
shown).  Hence, it is possible for errors, if they are big enough, to
create the \emph{impression} of planes by broadening a distribution
that is intrinsically merely a one-dimensional line.  For example, all
ellipticals might be the same age, obey the \enh--$\sigma$ relation (a
line), and be broadened by large \hbeta\ errors to fill apparent
``planes'' just like those observed.

The only defense against such an error is to know from independent
measurements that the observational errors are small.  That is why we
use only the G93 and Kuntschner (1998) samples, whose errors are small
and well understood.  The rms error of \hbeta\ in G93 is 0.060 \AA,
and in Kuntschner (1998) is 0.089 \AA, with errors in the other
indices being comparable.  As shown by the error ellipses in the
figures, these errors are small enough that the observed planes cannot
be artifacts.  Much larger errors, however, would be disastrous.  For
example, Figure~\ref{fig:tz} also shows the error ellipse for a
typical galaxy in the IDS sample of TWFBG98 ($\sigma_{\hbeta} = 0.191$
\AA for the 150 highest-quality galaxies).  Monte Carlo simulations of
this sample (\cite{Trager97}) have shown that the observed \z- and
Fe-planes were largely artifacts caused by observational errors; this
is consistent with the large size of the IDS error ellipse in
Figure~\ref{fig:tz}.  A reasonable guide is that \hbeta\ must be
accurate to $\sim0.1$ \AA\ to measure reliable ages and metallicities.

\subsection{Comparison with previous studies}\label{sec:sum_compare}

We compare next to other studies using Balmer-line data to determine
stellar population parameters.  The study by Kuntschner (1998) on
Fornax ellipticals is quite consistent with ours, which is not
surprising since we use the same data and similar models.
Kuntschner's conclusions were limited by the fact that his corrections
for non-solar abundance ratios were only approximate.  Nevertheless,
his findings that the Fornax ellipticals are mainly old and that they
show a strong \enh--$\sigma$ relation are confirmed here.

The study by J{\o}rgensen (1999) of 71 early-type galaxies in Coma is
similar in both approach and conclusions to the present work.
J{\o}rgensen (1999) analyzed newly obtained long-slit and multi-fiber
spectra and derived stellar population parameters using line-strength
models by Vazdekis et al.~(1996).  Overall her findings are similar to
ours, including a $\log\sigma$--[Mg/Fe] relation like that in
Figure~\ref{fig:sa}, an age--\mgh\ relation rather like that in
Figure~\ref{fig:tz}, and a tight age--\feh\ relation nearly identical
to that in Figure~\ref{fig:tf}.

However, the typical error of \hbeta\ in the J{\o}rgensen data is
$0.22$ \AA, with a long tail to larger errors.  Overall, her data are
comparable in accuracy to the IDS data of TWFBG98, which were found to
be inadequate for age determination by Trager (1997).  Given our
present understanding of the pernicious effects of errors
(Section~\ref{sec:errors}), we suspect that some of the trends found
by J{\o}rgensen are real but that others may be largely artifacts
caused by errors.  Specifically, the $\log\sigma$--[Mg/Fe] relation
found by J{\o}rgensen is almost certainly correct, whereas any of the
relations involving age (including both the \zp-plane and the
Fe-plane) are likely to be heavily contaminated.  A high-accuracy
line-strength survey of Coma ellipticals is badly needed.

The study of Tantalo, Chiosi \& Bressan (1998a) analyzed G93 data and
is thus relatively unaffected by observational errors.  A detailed
comparison to this work was made in Paper I.  The methodology of these
authors is very similar to ours except that only \mgb\ is corrected
for non-solar ratios whereas \fe\ is unchanged.  Their method
essentially measures \z\ based on \fe\ alone, and metallicities are
consequently underestimated and enhancements overestimated, by amounts
that increase with \enh.

Systematic errors increasing with \enh\ introduce slope errors into
most correlations.  For example, TCB98 find a strong \enh--age
relation, which seems to be our \enh--$\sigma$ relation lensed through
correlated errors.  The importance of this discussion is to show that
the factors used to correct line strengths for non-solar abundance
ratios---in particular the \emph{relative amplitudes} of the
corrections to \mgb\ and \fe---have far-reaching consequences for
parameter correlation studies.  Our corrections are based
self-consistently on the TB95 response functions, but independent
checks of those functions would be welcome.

\section{Other projections of the metallicity
hyperplane}\label{sec:other_views}

The notion that the stellar-population manifold of elliptical galaxies
is inherently two-dimensional is key to understanding many
two-parameter relationships involving these galaxies.  Most such
relationships are either projections of this higher-dimensional space
or are close relatives of such projections.  The slope and scatter of
points in such projections are not fundamental, but rather depend on
the distribution of points within the hyperplane.  The question of
sample selection thus enters acutely, as that may govern the
distribution of points in the plane.

\begin{figure*}
\plotonet{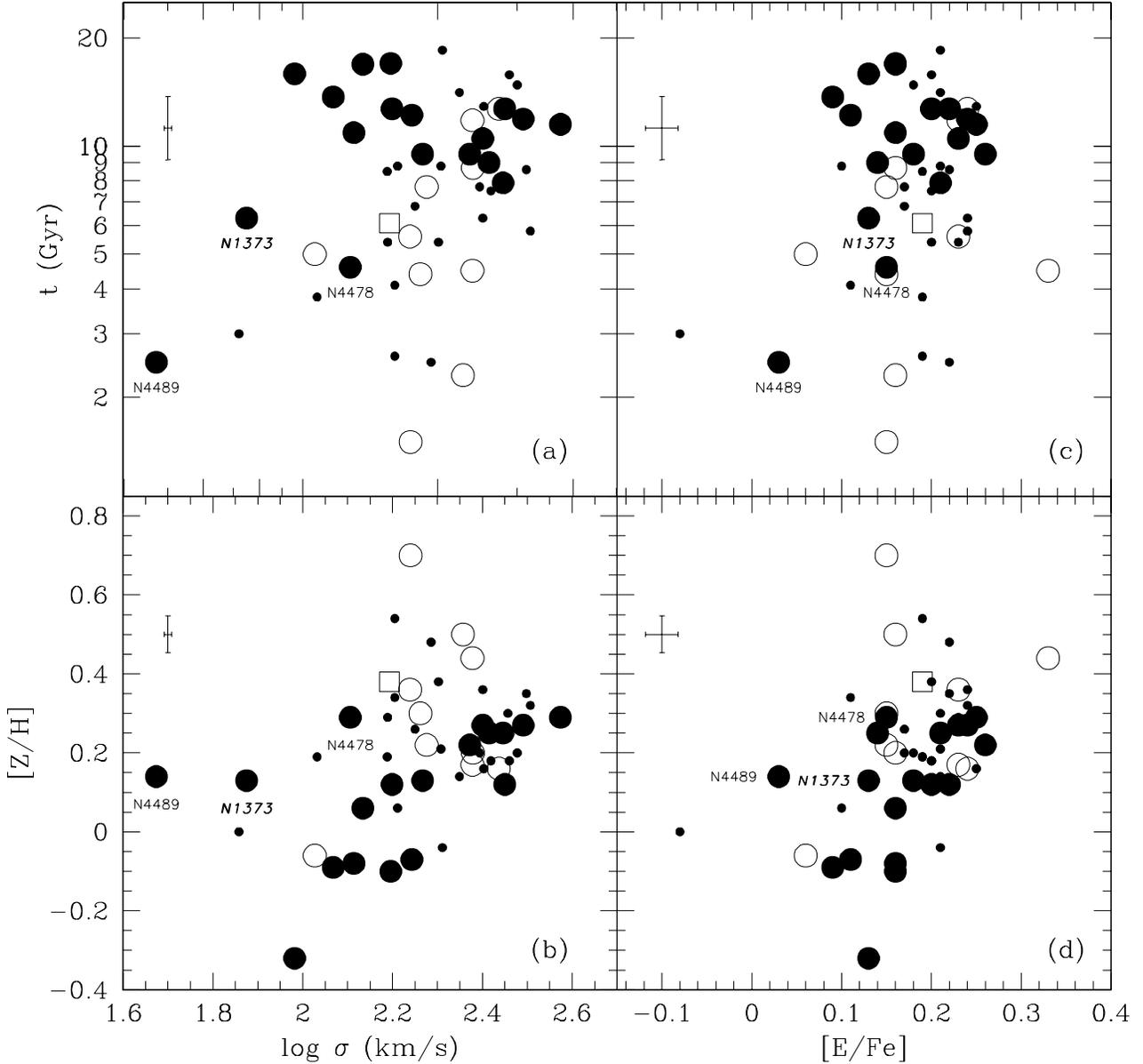}
\caption{Two-dimensional projections of the metallicity hyperplane,
coded by galaxy environment.  The figure illustrates how the
distribution of points in the \zp-plane affects projected
correlations.  Group assignments and group richnesses are taken from
Faber et al.~(1989) for most galaxies.  Large filled circles are
cluster galaxies (Virgo and Fornax); small filled circles are group
galaxies; open circles are isolated galaxies; and the large open
square represents the center of NGC 224 (M~31).  (a) The independent
variables $\sigma$--$t$, showing the different distributions of
various subsamples in the hyperplane.  Most cluster galaxies (large
filled circles) are old (with the notable exceptions of the small
galaxies NGC 4489, NGC 1373, and NGC 4478), whereas field galaxies
(groups $+$ isolated objects) span a large range in ages.  (b) The
$\sigma$--\z\ projection.  With the exception of the three outliers,
cluster galaxies trace a fairly well defined metallicity--$\sigma$
relation; field galaxies do not.  (c) The \enh--$t$ projection.  No
clear trends are seen in any subsamples.  (d) The \enh--\z\
projection.  A slight hint of an increase of \enh\ with \z\ is
apparent, but the scatter is large.\label{fig:tzas_proj}}
\end{figure*}

\subsection{The velocity dispersion--age projection}\label{sec:results_st}

Figure~\ref{fig:tzas_proj} shows several examples of how
two-dimensional projections are affected by the distribution of points
in the hyperplane.  Points are coded by the environment of each galaxy
in preparation for the discussion of environmental effects in the next
section.

Figure~\ref{fig:tzas_proj}a (upper left panel) shows the independent
variables $\sigma$ vs. $t$.  Since the $(\sigma,t)$ distribution
governs the appearance of all other projections, it is interesting to
compare the distributions within it of galaxy subsamples classed by
environment; isolated, group, and cluster ellipticals are shown by
open circles, small dots, and large dots respectively.  These
distributions look rather different; cluster E's (large filled
circles) are grouped near the top of the plot, except for three young
outliers shown by the labeled points: NGC~1373 is a bona fide member
of Fornax based on position and velocity yet is conspicuously young,
the only young Fornax elliptical; NGC~4489 is 4 degrees from the
center of Virgo but is a member by radial velocity.  It, too, is
rather young, as is NGC~4478, which is right near the center of Virgo
and is clearly a cluster member.  Within the errors, however, the bulk
of cluster galaxies is consistent with being old and coeval.

Group and isolated objects (which we collectively term ``field''
ellipticals; small dots and open circles) are distributed differently
from cluster ellipticals in the hyperplane.  They cover a larger age
range, and there is a weak trend in $t$ vs.~$\sigma$ in the sense that
low-$\sigma$ galaxies tend to be younger; the clump of old,
low-$\sigma$ galaxies that is prominent among the cluster galaxies is
also missing.

We conclude that the $(\sigma,t)$ distributions of local field and
cluster ellipticals differ in the present sample, and that their
two-parameter projections may also differ on that account.  That
prediction is explored in the following panels.

\subsection{The $\sigma$--\zp\ projection}\label{sec:results_sz}

Figure~\ref{fig:tzas_proj}b (lower left panel) plots \z\ vs. $\sigma$.
A velocity dispersion--metallicity relation appears to exist for old
cluster galaxies, but the three young cluster galaxies NGC 1373, NGC
4489, and NGC 4478 lie at higher \z\ at given $\sigma$.  No comparable
relation appears to exist for field ellipticals.  This difference is a
natural consequence of the differences in the $(\sigma,t)$
distributions above.  This projection is a close relative of the
classic mass-metallicity relation and is discussed further in
Section~\ref{sec:enveff}.

\subsection{The $t$--\enh\ projection}\label{sec:results_ta}

Figure~\ref{fig:tzas_proj}c (upper right panel) shows the distribution
of \enh\ as a function of age.  There is no apparent trend in \enh\
with $t$ in any sample.  This is as expected, since we found earlier
that \enh\ depends only on $\sigma$ and not on $t$.

\subsection{The \zp--\enh\ projection}\label{sec:results_za}

Figure~\ref{fig:tzas_proj}d (lower right panel) shows the distribution
of \enh\ as a function of metallicity \z.  There is a weak tendency
for \enh\ to increase with \z, especially among old cluster
ellipticals, suggesting higher SNe II/SNe Ia enhancement ratios in
higher-metallicity galaxies.  This trend among old cluster E's is
again expected from their narrow age distribution in
Figure~\ref{fig:tz}---in a narrow age range, \z\ increases with
increasing $\sigma$, and therefore \enh\ should increase with \z\ at
fixed $t$.

\subsection{Summary of results}\label{sec:sum_results}

Our results so far can be summarized as follows.  A principal
component analysis demonstrates that central stellar populations in
the present sample of local elliptical galaxies can be largely
specified using just two independent variables; we take these to be
SSP-equivalent age, $t$, and velocity dispersion, $\sigma$.  Velocity
dispersion is the only structural parameter that appears to play a
role in modulating the stellar populations of these galaxies.

This two-dimensional ``metallicity hyperplane'' is in turn comprised
of two sub-relations: metallicity is a linear function of both $t$ and
$\sigma$, which we call the ``\zp-plane,'' and enhancement ratio,
\enh, is a linear function of $\sigma$, increasing towards
high-$\sigma$ galaxies.  Together these two subrelations comprise the
hyperplane.

Several caveats are necessary.  First, the thickness of the hyperplane
appears to be at least partly real and is associated mainly with
scatter in the \enh--$\sigma$ relation.  Thus, the populations are not
perfectly two-dimensional, and at least one more factor must play a
role.  Second, the present SSP parameters are based on only three
spectral indices (\mgb, \fe, and \hbeta), and adding more indices (or
colors) might reveal more principal components; we will be
investigating this in future papers.  Third, coverage of the
hyperplane needs to be improved by adding more young populations,
which are relatively scarce here.  Fourth, the present data refer to
only 51 galaxies; a larger sample is needed to confirm that the
present trends in fact apply to local elliptical galaxies generally.
Fifth, we must remember that the hyperplane refers to
\emph{SSP-equivalent} population parameters, which are
disproportionately influenced by young stars (see Appendix).  However,
despite the fact that SSP-parameters are not true mass-weighted
averages, they still place very tight constraints on the history of
star formation in ellipticals, as shown below in
Sections~\ref{sec:disc_sz} and \ref{sec:disc_plane}.

Finally, the present analysis delineates the position and orientation
of the hyperplane in hyperspace but says little about the distribution
of galaxies within it.  That is because our sample does not constitute
an unbiased \emph{volume-limited} sample of local ellipticals.  This
places severe limits on our conclusions. For example, we cannot
conclude that the wide range of of SSP ages seen in our field galaxies
is typical of local field ellipticals generally.  However, there is a
strong suggestion in our data that the $(\sigma,t)$ distributions of
field and cluster galaxies may differ, with cluster ellipticals in the
present sample being generally older.  This difference is expected to
generate environmental differences in the two-dimensional projected
scaling laws of these galaxies, as explored in the next section.

\section{Two classical scaling laws}
\label{sec:scaling_laws}

This section investigates two classical scaling laws for elliptical
galaxies: the mass-metallicity relation and the Mg--$\sigma$ relation.
Both can be understood as two-dimensional projections of the
metallicity hyperplane.

\subsection{The mass-metallicity relation: environmental effects}
\label{sec:enveff}

Environmental differences among elliptical galaxies have generated
intense interest (e.g., \cite{dCD92}; \cite{BFD90}; \cite{Guzman92};
\cite{Bernardi98}).  We consider here their impact on a question of
major importance, the mass-metallicity relation of elliptical
galaxies, widely regarded as a key clue to their nucleosynthetic
histories (e.g., \cite{AM85}).  The relation comes in several guises:
\z\ vs.\ mass, \z\ vs.\ luminosity, and \z\ vs.\ $\sigma$---this last
also counts as a mass-metallicity relation since mass and $\sigma$ are
so closely correlated.

\begin{figure*}
\plotonet{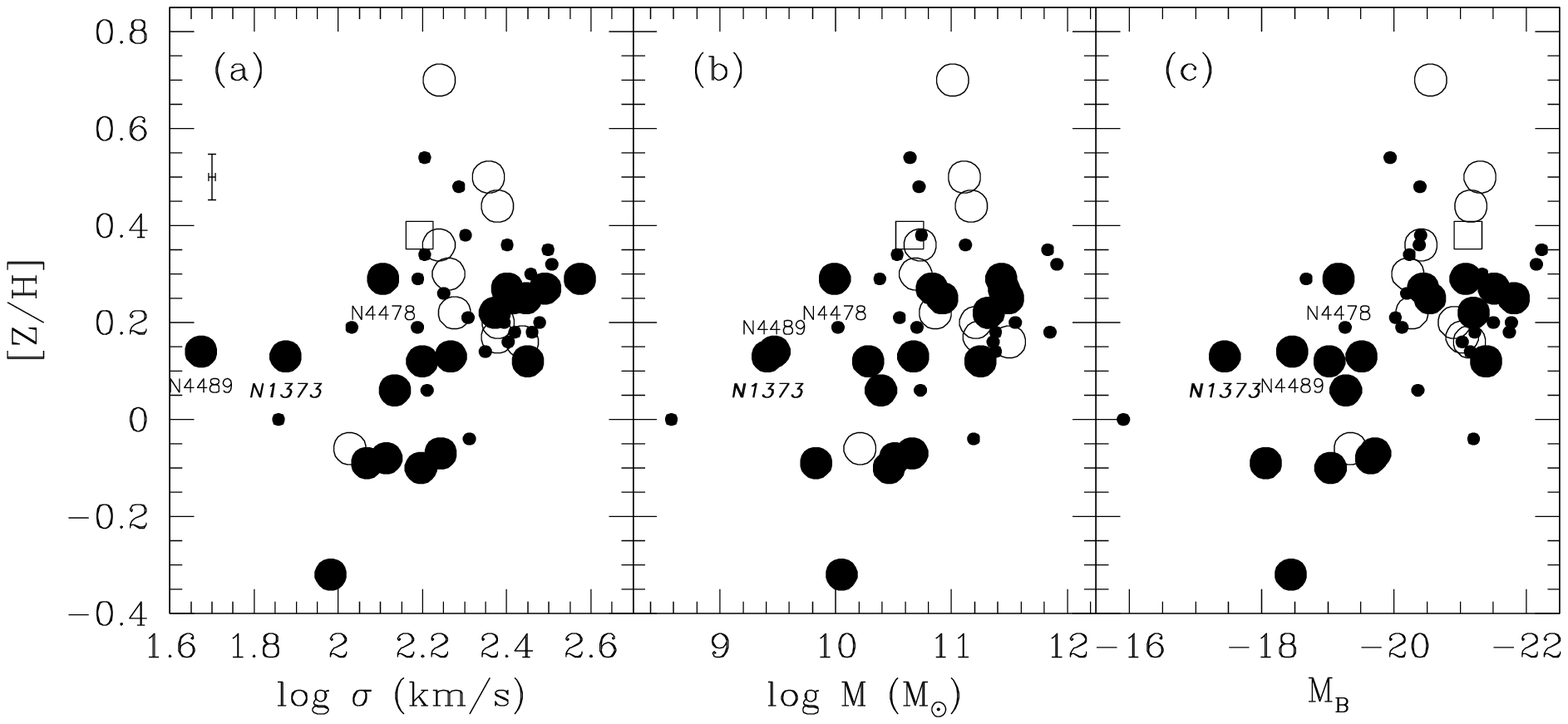}
\caption{Three ``mass''--metallicity relations for the centers of
local ellipticals, as a function of environment.  (a) The velocity
dispersion--metallicity projection, reproduced from
Figure~\ref{fig:tzas_proj}b.  (b) The actual mass--metallicity
relation.  (c) The luminosity--metallicity relation, which closely
resembles the mass-metallicity relation in panel (b).  A weak trend
with galaxy size may be apparent in all three panels, but the scatter
is large, and the precise relations may be sample dependent.  The
distributions of points in these diagrams are determined by the sample
distributions in the $(\sigma,t)$ plane
(cf. Figure~\ref{fig:tzas_proj}).
\label{fig:massz}}
\end{figure*}

These three projections are compared in Figure~\ref{fig:massz}.  The
\z--$\sigma$ projection (panel a) is repeated here from
Figure~\ref{fig:tzas_proj}.  We have already observed that any
relation in this panel is weak; old cluster galaxies (large filled
circles) show a trend in the classic sense that high-$\sigma$ galaxies
are more metal rich, but this trend is not shared by field galaxies
(open circles and small dots).  Panels b and c show \z\ vs. mass and
\z\ vs. absolute magnitude (the latter quantities are taken from
Table~\ref{tbl:ddq}).  These relations show even more scatter than \z\
vs. $\sigma$, and the real mass-metallicity relation (panel b) is
worst of all.

It is not our purpose to argue here that there is \emph{no}
mass-metallicity relation.  Rather, like many two-dimensional
correlations claimed for elliptical galaxies, the mass-metallicity
relation is actually a projection of a higher-dimensional space.  As
such, it may be both environmentally and sample dependent, and its
accurate determination will require a larger and more carefully
controlled sample than we have here.

\subsection{The \mg--$\sigma$ relations}\label{sec:mgsigma}

The Mg--$\sigma$ relations present a major challenge to the hyperplane
model.  The tightness of these relations has often been taken as
evidence that all ellipticals have nearly coeval stellar populations
to of order 15\% in age (\cite{BBF2}; \cite{Bernardi98}), in strong
contradiction to the spread of about a factor of 10 in SSP ages found
in this work.  We are planning a separate paper on this important
issue but include a short section here in order to address pressing
questions that will occur to knowledgeable readers.

Our picture is that the Mg--$\sigma$ relations look narrow because
they are edge-on (or nearly edge-on) projections of the metallicity
hyperplane.  The germ of this idea is contained in
Figure~\ref{fig:tz}, which shows the \zp-plane face on.  Imagine
rotating this plane about an axis running perpendicular to the
contours of constant $\sigma$ and viewing the resultant projection
edge-on.  Suppose further that SSP-equivalent age and metallicity
``conspire'' to cause \mgb\ (or \mgtwo) to remain sensibly constant
along a $\sigma$--contour.  This would occur if
$\Delta\logt/\Delta\z=-1.7$ or $-1.8$ (W94), and indeed the \zp-plane
at fixed $\sigma$ (Equation~\ref{eq:tzs}) has slope very close to
this: $\Delta\logt/\Delta\z=-1.4$.  In other words, lines of constant
$\sigma$ closely obey the 3/2 rule, and line-strength along them
should be nearly constant.  In projection, \mgb\ and \mgtwo\ should
therefore be tight functions of $\sigma$, yielding the Mg--$\sigma$
relations.

\begin{figure*}
\plotonet{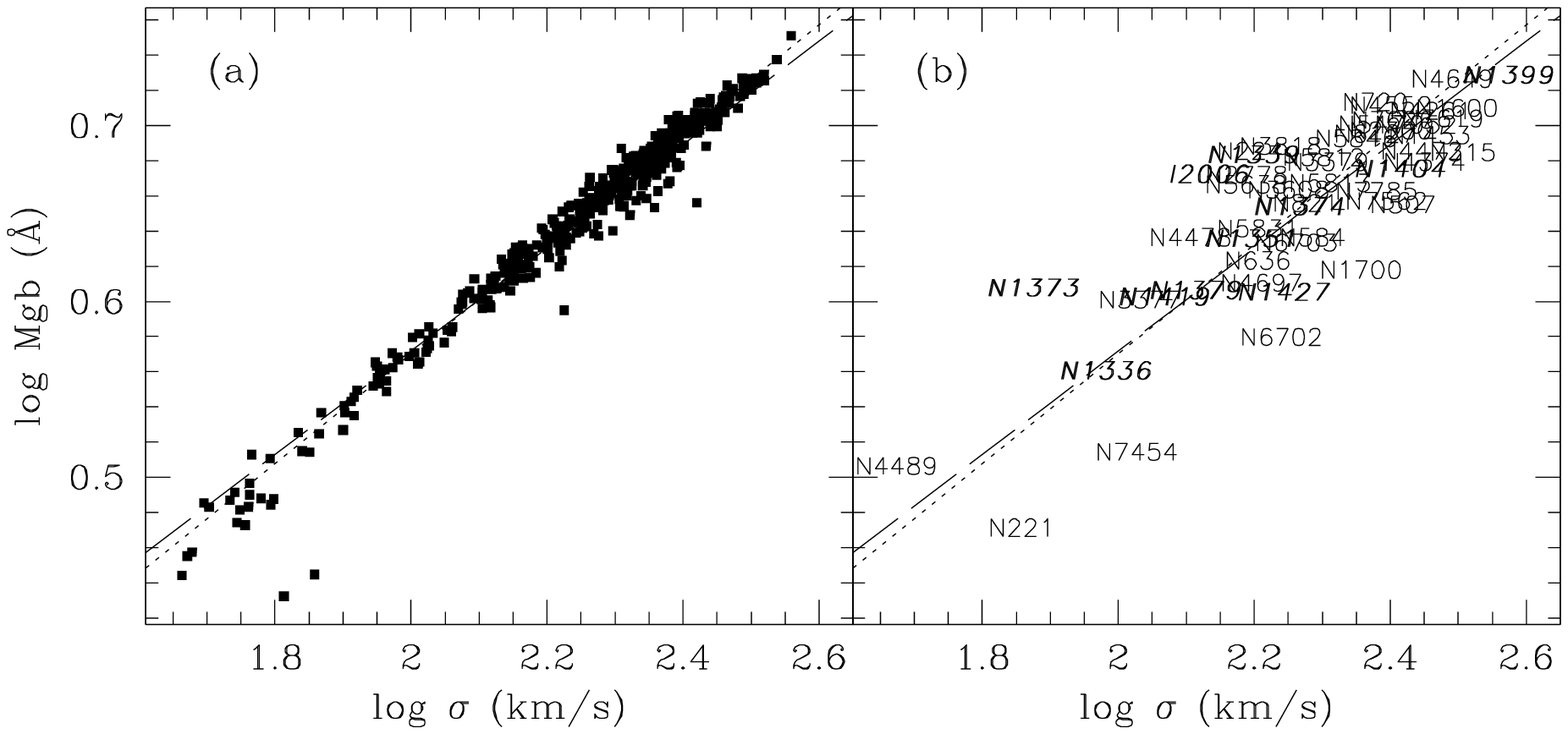}
\caption{(a) Simulated \mgb--$\sigma$ relation based on an assumed
infinitely thin hyperplane.  The dotted line is a least squares fit of
the form $\log\mgb=0.312\log\sigma-0.054$, with an rms scatter of only
0.007 (485 realizations).  This is virtually the same as the dashed
line from panel b showing the actual regression line of the present
sample.  (b) The actual \mgb--$\sigma$ relation of the present sample.
The dashed line is a least squares fit of the form
$\log\mgb=0.294\log\sigma-0.016$, with an rms scatter of 0.032 (51
galaxies).
\label{fig:mgbsigma}}
\end{figure*}

To illustrate this graphically, we have performed Monte Carlo
simulations to produce \mgb\ and $\sigma$ values for roughly 500
``fake'' elliptical galaxies realistically distributed in the
metallicity hyperplane.  Random values of the first two principal
components in Table~\ref{tbl:pca} were drawn from the distribution of
galaxies in the face-on view of the plane (Figure~\ref{fig:pca}), and
the third and fourth components were set identically to zero.  These
four PCA eigenvectors were then inverted to determine $t$, \z, \enh,
and $\sigma$ for each realization, and the first three parameters were
used to generate line strengths using the formalism described in Paper
I, with typical observational errors added.  The resulting simulated
\mgb--$\sigma$ relation is shown in Figure~\ref{fig:mgbsigma}a.  The
derived relation (dotted line) has the form
\begin{eqnarray}
\log\mgb &=&\phm{\pm}0.312\,\log\sigma - 0.054,\\
         & &\pm0.002\phm{\,\log\sigma}\pm0.001\nonumber
\end{eqnarray}
with an RMS scatter of only 0.007.  The \mgb--$\sigma$ relation for
the present sample of local elliptical galaxies (dashed line, panel b)
has the form
\begin{eqnarray}
\log\mgb &=&\phm{\pm}0.294\,\log\sigma - 0.016,\\
         & &\pm0.005\phm{\,\log\sigma}\pm0.001\nonumber
\end{eqnarray}
with an RMS scatter of 0.032 (51 galaxies).  The good agreement
between the simulated relation and the real one confirms that a large
age spread of stellar populations in the hyperplane can indeed be
masked by the tendency of \z\ to rise at low ages, precisely
compensating the effect of age differences.

We briefly mention a few important points, saving details for
our future paper:

(1) The idea that the tightness of the Mg--$\sigma$ relations might
conceal large age variations was first proposed by Worthey et
al. (1996) and was re-proposed by J{\o}rgensen (1999). In both cases,
the actual age spreads were probably somewhat overestimated, as
Worthey et al.\ used the Lick/IDS data while J{\o}rgensen used her
Coma sample, both of which contain significant observational errors
(Sections~\ref{sec:errors} and \ref{sec:sum_compare}).  Nevertheless,
the basic correctness of the idea is confirmed here.

(2) While the slopes of the real and simulated \mgb--$\sigma$
relations match well, the scatter in the simulated relation is too
small even though observational errors have been included.  That is
because the third and fourth principal components were neglected,
i.e., the hyperplane was taken to be infinitely thin.  This was done
deliberately to make any residual tilt of the hyperplane more visible.
Even with this, the simulated relation is still extremely narrow,
showing that any deviation from an edge-on orientation must be small.
The larger scatter of the real \mgb--$\sigma$ relation must be due to
the presence of PC3 and PC4, which were not included in the
simulation.  PC3, in particular, reflects real scatter in the
\enh--$\sigma$ relation, as noted in Section~\ref{sec:results_sa}.

(3) Although the Mg--$\sigma$ relations are generally tight,
morphologically disturbed ellipticals tend to show lower Mg values
than expected, and this has been convincingly interpreted as due to
recent star formation by Schweizer et al. (1990) and Schweizer \&
Seitzer (1992). Comparably young stellar populations are present in
some of our galaxies here (e.g., NGC 6702, NGC 5831, NGC 1700), yet
none of these shows any significant deviation from \mgb--$\sigma$ in
Figure~\ref{fig:mgbsigma}.  Is this a disagreement?

A full discussion of this point is reserved to our future paper, but
we can sketch the answer briefly here.  First, the Mg relations used
by Schweizer et al. (1990) and Schweizer \& Seitzer (1992) actually
plotted Mg vs. luminosity, $L$, not $\sigma$.  Recent star formation
would increase $L$ while depressing Mg, thus amplifying any Mg
residual.  Second, a handful of low-lying points can be seen in the
simulated \mgb--$\sigma$ relation in Figure~\ref{fig:mgbsigma}.  These
turn out to be the youngest galaxies, demonstrating that slight
curvature in the transformations back to raw \mgb\ can cause objects
to lie low if they are extremely young.  Finally, essentially all
previous investigations of Mg--$\sigma$ have used \mgtwo, whereas we
chose \mgb\ because it was more accurately measured by G93.  This
decision proves to be important, as separate other work now shows that
\mgtwo--$\sigma$ is not as tight as \mgb--$\sigma$ and does indeed
show small but systematic negative residuals for younger stellar
populations.  This is evident both in the present sample and in the
larger Lick/IDS sample of TWFBG98.

Thus, it appears that both views are correct: the basic tightness of
the Mg--$\sigma$ relations conceals large age spreads, but \mgtwo\ in
particular deviates systematically in the sense that young stellar
populations lie low.  Further discussion of these and other aspects of
the Mg--$\sigma$ relations will be provided in our future paper.

\section{The origin of the \enh--$\sigma$ relation}
\label{sec:disc_sz}

We have seen that there are two major correlations involving the
stellar populations of the present sample: the \zp-plane linking \z,
$t$, and $\sigma$; and the \enh--$\sigma$ relation linking \enh\ and
$\sigma$.  Assuming that these relations are in fact a good
description of local ellipticals generally, we attempt to deduce the
implications for their star formation histories.  To anticipate, we
find a number of plausible explanations for \enh--$\sigma$; the
relation is interesting and useful but in retrospect not very
surprising.  The existence of the \zp-plane on the other hand turns
out to be very puzzling and may emerge as one of the most telling
constraints on the history of star formation in ellipticals.  This
section focuses on the simpler \enh--$\sigma$ relation; theories for
the \zp-plane are explored in the next section.

Six scenarios for \enh--$\sigma$ are considered; findings are
summarized as a truth table in Table~\ref{tbl:truthtable}.  Each
scenario is compared to three observed trends in a binary, yes-no
way---does the scenario account for the observed trend or not?  The
first trend is the \enh--$\sigma$ relation itself, which is given
highest weight.  We also add two additional ``trends,'' that \z\ and
\feh\ both increase with $\sigma$.  These trends are true strictly
speaking only at fixed $t$ (Equations 8 and 10), and thus apply only
to populations with a narrow range of SSP ages, e.g., cluster
galaxies.  Since all ellipticals are clearly \emph{not} the same SSP
age, using these extra trends may be unwarranted.  However, adding
them narrows the possibilities greatly, and it is perhaps reasonable
to require that any successful scenario for the \enh--$\sigma$
relation must separately explain old cluster galaxies.  Most of the
ideas below have been discussed in the literature before, but the
present information on \enh, \z, and \feh\ separately sheds new light.

Strictly speaking, our measurements refer to SSP values of \enh, which
are heavily weighted by young stars.  However, experiments in
Section~\ref{sec:disc_plane} suggest that mixed-age ``frosting''
models must have rather constant values of \enh\ in all
sub-populations in order for composite galaxies to match the
\zp-plane.  In such cases, SSP values of \enh\ are a good
mass-weighted mean for the whole population.

\begin{deluxetable}{lccc}
\tablecaption{Scenarios for the \enh--$\sigma$ relation\label{tbl:truthtable}}
\tablewidth{0pt}
\tablehead{&&\colhead{$\z\uparrow$ with $\sigma\uparrow$}
&\colhead{$\feh\uparrow$ with $\sigma\uparrow$}\\
\colhead{Scenario}&\colhead{$\enh\uparrow$ with $\sigma\uparrow$?}&
\colhead{(at fixed $t$)?}&\colhead{(at fixed $t$?)}}
\tablecolumns{4}
\startdata
1) No. stellar generations increases  as $\sigma$ increases&n&y&y\cr
2) Star formation duration decreases as $\sigma$ increases&y&n&n\cr
3) Late winds reduce SN Ia yield as $\sigma$ increases&y&n&n\cr
4) Number of Type Ia SNe decreases as $\sigma$ increases&y&n&n\cr
5) IMF flattens as $\sigma$ increases&y&y&y\cr
6) Early winds reduce SN II yield as $\sigma$ decreases&y&y&y\cr
\enddata
\end{deluxetable}

The scenarios are as follows:

(1) \emph{The number of stellar generations (i.e., total astration)
increases with increasing $\sigma$.}  This scenario has roots in the
classic closed box model and envisions that star-formation and cosmic
recycling go further at higher $\sigma$ (assuming that the relative
yields from Type Ia and Type II SNe do not change).  This scenario can
account for higher \z\ and \feh\ with higher $\sigma$ but clearly does
not predict any change in \enh.  It is included for completeness only.

(2) \emph{The duration of star formation is shorter with increasing
$\sigma$.}  This scenario envisions that the total duration of star
formation (in years, not in stellar generations) is reduced at high
$\sigma$ (e.g., \cite{WFG92}).  Such shortening would reduce the
amount of Fe--peak elements because star formation would be over
before SNe Ia exploded and their Fe--peak products became available
for incorporation into new stars.  In this scenario, total astration
through SNe II remains the same, but elements from SNe Ia are reduced.
This matches the observed increase in \enh\ with $\sigma$, but,
because total element production is also reduced, it cannot match
either the increase in \z\ or \feh\ with higher $\sigma$.

Scenarios 1 and 2 were designed to separate the notion of the
\emph{number of generations} of element building (astration) from the
\emph{number of years} needed to form those generations (duration).
Since the two scenarios have complementary failings in
Table~\ref{tbl:truthtable}, one may wonder whether combining them
(shorter formation time plus more astration at high $\sigma$) might
match all the data.  This is a quantitative question whose answer
depends on detailed model parameters and calculations.  Our impression
is that such a model could likely match the increase in \enh\ and \z\
with $\sigma$ but would probably have flat or falling \feh\
vs. $\sigma$, contrary to the data.  Even more difficult is the fact
that, in nature, astration and duration are naturally positively
coupled---longer star formation means there is time for more
astration---not anti-coupled as in this hybrid model.  Such coupling
is seen, for example, in the models of Larson (1974), Arimoto \&
Yoshii (1987), and Thomas, Greggio \& Bender (1999).  Moreover, in all
these cases, as star formation proceeds, recycling of material through
SNe Ia causes \enh\ to decrease and metallicity and \feh\ to rise.
\enh\ is therefore naturally \emph{anti}-correlated with the others,
unlike the data.  For both reasons, combining scenarios 1 and 2 does
not seem promising.

(3) \emph{Late winds are stronger with increasing $\sigma$.}  This
scenario is essentially a carbon-copy of scenario 2 in that both serve
to reduce the amount of SN Ia-enriched material retained by the galaxy
while leaving SN II products unchanged.  Like scenario 2, it matches
the increase in \enh\ with $\sigma$ but predicts a fall in both \z\
and \feh\ at high $\sigma$, contrary to observations. Moreover, it is
inherently implausible that galactic outflows should be \emph{higher}
in high-$\sigma$ galaxies, which have deeper potential wells.

(4) \emph{The number of Type Ia SNe decreases with increasing
$\sigma$.}  If SNe Ia are explosions of double-degenerate systems as
is generally assumed (e.g., \cite{WH90}), their progenitors are tight
binaries.  It may be that, in a high-$\sigma$ environment, glancing
cloud-cloud collisions impart enough angular momentum to form only
very wide binaries, and thus suppress the formation of SNe Ia
progenitors.  Intriguing as this speculation is, the net result of
this proposal is again not very different from the previous two
scenarios, which reduce elements from SNe Ia while leaving those from
SNe II unchanged.  It fails for the same reasons.

The next two scenarios \emph{increase} element yields from SNe II
while leaving leaving those from SNe Ia unchanged.  These are more
successful.

(5) \emph{IMF flattens with increasing $\sigma$.}  In this scenario,
more high-mass stars are born and more SNe II are produced in
high-$\sigma$ galaxies, increasing the effective yield and thus the
overall mean metal abundance of the stellar population
(\cite{Tinsley80}).  The quantitites \enh, \z, and \feh\ all increase
with $\sigma$ (this last because Type II SNe produce at least some Fe;
\cite{WW95}).  However, the increase in \feh\ should be weaker than in
\z, as is observed (compare Equations~\ref{eq:tzs} and~\ref{eq:tfs}).
Although this scenario matches all the data, no physical mechanism for
it is as yet known.  Perhaps massive star-formation is enhanced at
high cloud-cloud collision velocities, which in turn would scale in
rough proportion to stellar velocity dispersion (\cite{FWG92}).
\footnote{This is the place to clarify a potentially confusing aspect
of our terminology.  Earlier we stressed that high values of \enh\ do
not reflect an ``enhancement'' of the E elements but rather a
depression of the Fe--peak elements, yet here scenario 5 accounts for
high \enh\ by ``increasing'' the effective yield of Type II elements.
We seem to be saying simultaneously that the E elements are enhanced
and not enhanced.  Actually, these two statements are not in
contradiction.  The non-enhancement mentioned earlier refers to [E/Z],
which is always near zero since E effectively \emph{is} Z.  Scenario 5
deals on the other hand with \enh, which clearly \emph{can} be
increased by raising the absolute yield of Type II elements over Type
Ia.  The quantities [E/Z] and the yield of the E elements are not the
same, and one can be ``enhanced'' and not the other.}

(6) \emph{Early winds are stronger with decreasing $\sigma$.}  In this
scenario, all ellipticals produce SN Ia and SN II products at the same
rate, but low-$\sigma$ galaxies lose their early, SN II-enriched gas
more readily than high-$\sigma$ galaxies (see \cite{Vader86}, 1987 for
an early discussion of this process).  High-$\sigma$ galaxies would
have a higher effective yield of Type II SNe products, resulting in a
positive \enh--$\sigma$ relation and, because of their higher
retention of Type II SNe products, higher overall metallicities as
well.  Since Type II SNe make \emph{some} Fe (see above), \feh\ should
also increase weakly with $\sigma$, as is seen.  Observationally,
abundance trends in this scenario are similar to those of scenario 5,
in which the IMF is modulated by $\sigma$.

From hydrodynamic simulations of the mechanical effects of
supernovae-driven superbubbles on the gas and metal content of dwarf
galaxies, Mac Low \& Ferrara (1998) have shown that moderate starburst
events (SN II rates of $>3\;\mathrm{Myr^{-1}}$) in even massive dwarf
galaxies ($10^9\;M_{\odot}$) can blow out a substantial fraction
($\sim70\%$) of metal-enriched gas without losing a significant amount
of primordial gas ($<0.001\%$).  This process might be more important
for SNe II, which are highly spatially and temporally correlated, than
for SNe Ia, which seem to be relatively isolated in both time and
position within a galaxy.  This may enable low-$\sigma$ galaxies to
lose their SN II products preferentially without losing gas that can
later be enriched by SNe Ia and recycled into new stars.

Although scenarios 5 and 6 predict similar abundance \emph{trends}
with $\sigma$, they appear to differ in their absolute abundance
ratios.  With ``normal'' yields, the early winds in scenario 6 would
result in lower-than-normal abundances of Type II products in
low-$\sigma$ galaxies, but normal abundances in high-$\sigma$
galaxies, where all products are retained.  This is not as observed;
\enh\ is solar in low-$\sigma$ galaxies and enhanced in high-$\sigma$
galaxies (Figure~\ref{fig:sa}).  To work, scenario 6 may therefore
have to be ``tweaked'' by a blanket upward adjustment of the Type II
yield in \emph{all} elliptical galaxies, designed to return \enh\ in
low-$\sigma$ galaxies to the solar value.  Such a tweak might be
achived, for example, by boosting the upper end of the IMF in all
ellipticals by a similar amount.  This requirement would constitute an
additional burden on scenario 6.

In summary, there appear to be two viable scenarios that can currently
account for all three observational trends with $\sigma$: (1) a
flatter top end of the IMF that produces more massive stars at high
$\sigma$, and (2) weaker early winds, less mass loss, and greater
retention of SN II products at higher $\sigma$.  Although we cannot
tell which hypothesis is better, it is interesting, and a significant
step forward, that the data seem to prefer scenarios in which it is
the number or effectiveness of Type II SNe that are modulated, not the
number of Type Ia's.  A further new clue is that \enh\ correlates most
tightly with $\sigma$ and not with other related structural
parameters, such as mass or radius.  This tells us that the processes
modulating Type II SNe depend directly on the actual speeds of gas
clouds, or possibly on the escape velocity from the galaxy.  Finally,
it is necessary to restate the disclaimer that to reach these firmer
conclusions required using all three observational tests, including
the two less universal correlations involving \z\ and \feh.  If these
were thrown out, five out of the six scenarios would still be viable.

\section{The origin of the \zp-plane}\label{sec:disc_plane}

The origin of the \zp-plane proves to be more telling and more
difficult to explain than the \enh--$\sigma$ relation.  Two basic
star-formation scenarios for ellipticals are considered: (1) a pure
single-burst population having the measured SSP age and composition,
and (2) a double-burst population consisting of an old ``base''
population with a ``frosting'' of young stars.  More complex scenarios
can be inferred by extrapolating the results of the two-burst model.

\subsection{Single-burst stellar populations and their
evolution}\label{sec:pure_ssp}

\begin{figure*}
\plotonet{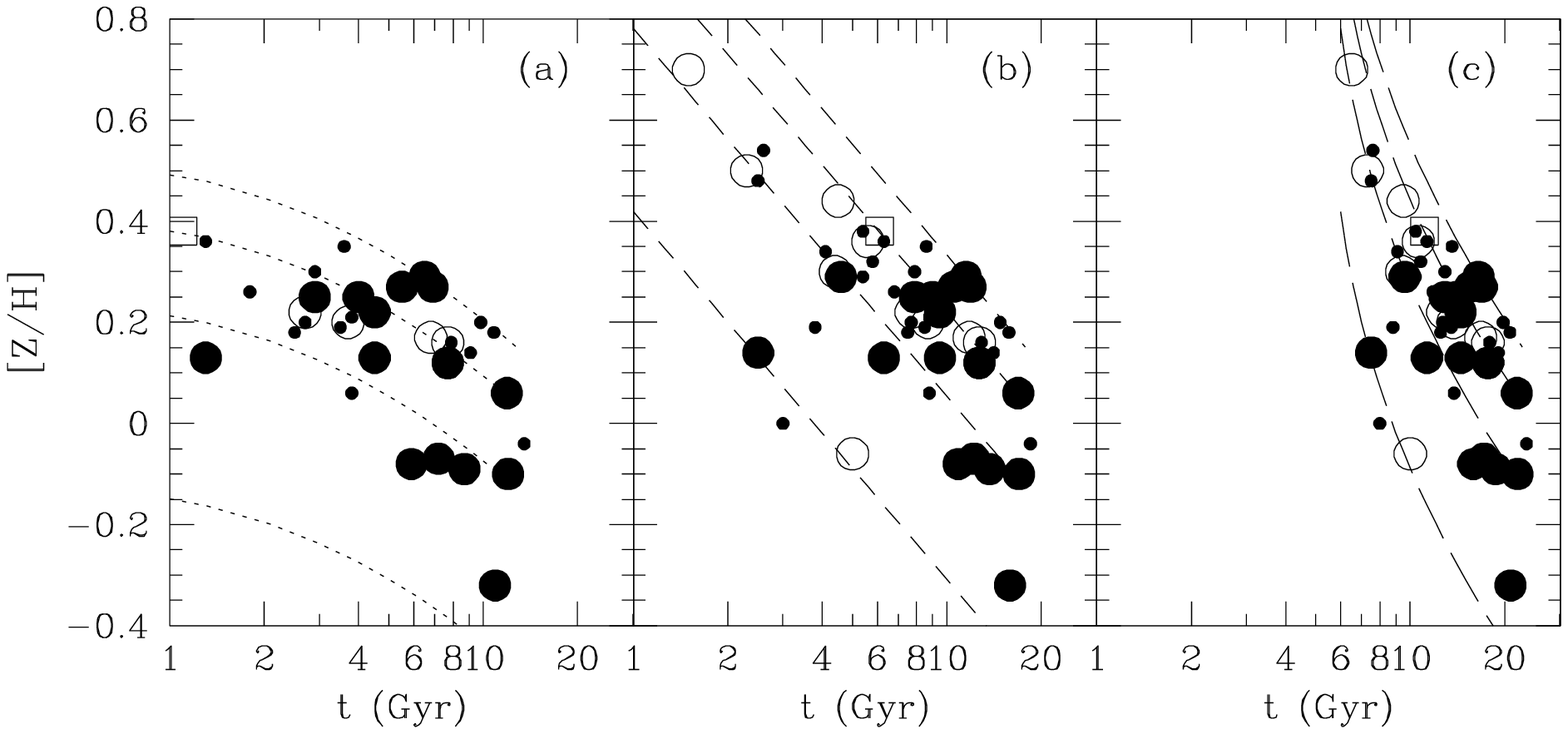}
\caption{The time evolution of galaxies in the \zp-plane for pure
SSPs.  Points are coded by environment (see
Figure~\ref{fig:tzas_proj}).  Lines are loci of constant velocity
dispersion: from bottom to top, $\sigma_0=50$, 150, 250, 350 \kms\
(see Figure~\ref{fig:tz}).  (a) The metallicity hyperplane 5 Gyr ago.
(b) The metallicity hyperplane today.  (c) The metallicity hyperplane
5 Gyr from now.  Note the strong curvature in lines of constant
$\sigma$ in panels (a) and (c).\label{fig:tz_evolve}}
\end{figure*}

Under the single-burst hypothesis, we observe that the \zp-plane is in
place at the present time (Figure~\ref{fig:tz}) and ask how it evolved
in the past and how it will evolve in the future.  The evolution of
the \zp-plane under pure single-burst SSP populations is simple:
galaxies move horizontally in $t$ as they age but stay constant in
both \z\ and $\sigma$.  Figure~\ref{fig:tz_evolve} shows this
behavior.  Note that since the ordinate is $\logt$ and not (linear)
$t$, old objects move less per unit time today than young objects.
Lines of constant $\sigma$ therefore \emph{steepen} into the future,
and after enough time they actually curve upwards.  This curvature
becomes pronounced after 5 Gyr, as shown in
Figure~\ref{fig:tz_evolve}c.  Similarly, lines of constant $\sigma$
curve \emph{downwards} in the past, as seen in
Figure~\ref{fig:tz_evolve}a.  Under the single-burst hypothesis, we
must therefore live at the special time when the \z--$t$--$\sigma$
surface is planar---i.e., lines of constant $\sigma$ are straight only
at the present time.  This seems improbable.

There are two additional problems with the single-burst scenario.  In
the rather recent past, many young galaxies seen today would not exist
at all if their populations really are pure SSPs.  For example, 12 of
51 galaxies (24\%) in the present sample would not have existed just 5
Gyrs ago (note how they have disappeared from
Figure~\ref{fig:tz_evolve}a).  Second, if the monotonically rising
age--metallicity relation at constant $\sigma$ that is seen today is
not special to this moment but will persist in future, the
metallicities of newly formed young galaxies must be rising very
rapidly at the present time.  In a few Gyr from now, new populations
will have to have metallicities in excess of $\z\sim+1$ (ten times
solar)!  Both of these problems illustrate again that the \zp-plane is
a short-lived, ephemeral phenomenon under the single-burst hypothesis,
and that our present epoch would have to be very special.

\subsection{Frosting models and their evolution}\label{sec:frosting}

The second scenario is the simplest composite stellar population
model, a double-starburst model in which a small ``frosting'' of young
stars forms on top of an old, ``base'' population. Examples of such
models and their behavior are discussed in the Appendix.  To a first
approximation, SSPs add vectorially (when weighted by light) in the
\hbeta--\mgb\ and \hbeta--\fe\ diagrams if the populations are not
very far apart, but trajectories between widely separated populations
are curved and must be calculated explicitly.  We do this by computing
light-weighted mean values of \hbeta, \mgb, and \fe, from which the
SSP-equivalent parameters are computed using the formalism described
in Paper I.

\begin{figure*}
\plotonet{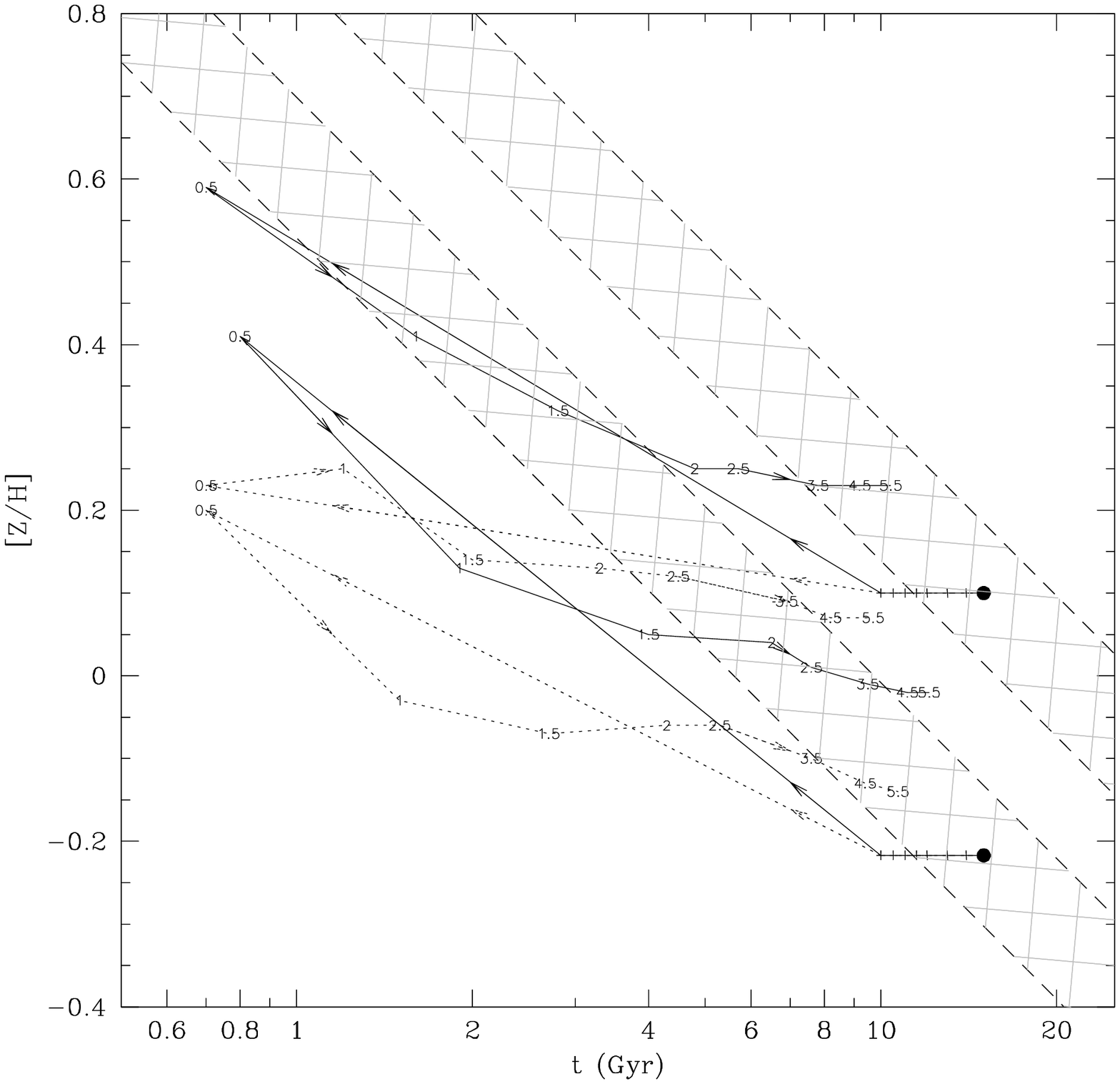}
\caption{The time evolution of frosting models in the $(t,\z)$
projection of the \zp-plane.  Two base models are shown: a ``giant''
elliptical base population ($\sigma=250\,\kms$,
$\z_{\mathrm{base}}=0.10$ dex, and $\enh_{\mathrm{base}}=0.22$ dex)
and a ``dwarf'' elliptical base population ($\sigma=100\,\kms$,
$\z_{\mathrm{base}}=-0.22$ dex, and $\enh_{\mathrm{base}}=0.08$ dex);
\z\ and \enh\ were chosen to place the base populations on the
present-day metallicity hyperplane at $t=15$ Gyr (large filled
circles).  SSP-equivalent populations that today lie in the
gray-hatched bands should have the same $\sigma$ as the base
population to lie in the observed \zp-plane (the width of the bands is
the typical $\pm1\sigma$ uncertainty in \z).  Two frosting populations
are shown for each base population: a solar-metallicity frosting
($\z_{\mathrm{frost}}=0.0$ dex and $\enh_{\mathrm{frost}}=0.0$ dex;
dotted trajectories) and a metal-rich frosting
($\z_{\mathrm{frost}}=0.50$ dex, $\enh_{\mathrm{frost}}=0.25$ dex;
solid trajectories).  Each frosting is 20\% by mass of the total
population and turns on at 9.5 Gyr.  The bursts are allowed to age for
5.5 Gyr until a final age of the base population of 15 Gyr.  The
crosses represent the passive evolution of the base population as seen
at 10, 10.5, 11, 11.5, 12, 13, 14, and 15 Gyr.  The composite
population is marked with a number representing the time in Gyr after
the starburst of the frosting population.  (Frosting populations at
0.5 Gyr were generated from the Padua models of Appendix A of Paper
I, for which line strengths are available down to 0.4 Gyr.)  After
aging for $\sim3$-5 Gyr, the composite populations successfully pass
through the same band of constant $\sigma$ as the base population (but
at higher SSP metallicity and younger SSP age) only if
$\z_{\mathrm{frost}}$ exceeds $\z_{\mathrm{base}}$ by 0.1 to 0.6 dex
and $\enh_{\mathrm{frost}} \sim \enh_{\mathrm{base}}$.
\label{fig:burst_evolve}}
\end{figure*}

\begin{deluxetable}{lrrrrrrrrrrrr}
\tablefontsize{\footnotesize}
\tablecaption{Evolution of two-burst frosting
models\label{tbl:burst_evolve}}
\tablewidth{0pt}
\tablehead{
&\multicolumn{3}{c}{Base}&
\multicolumn{3}{c}{Frosting}&
\multicolumn{6}{c}{Composite} \\
\colhead{Model\tablenotemark{a}}&
\colhead{$t$}&\colhead{\z}&\colhead{\enh}&
\colhead{$t$}&\colhead{\z}&\colhead{\enh}&
\colhead{\hbeta}&\colhead{\mgb}&\colhead{\fe}&
\colhead{$t$}&\colhead{\z}&\colhead{\enh}}
\tablecolumns{13}
\startdata
GS&10.0&$0.10$&0.21&0.5&0.00&0.00&4.19&2.66&2.07&0.7&$0.23$&0.26\nl
&10.5&&&1.0&&&2.67&3.64&2.50&1.2&$0.25$&0.22\nl
&11.0&&&1.5&&&2.13&4.16&2.65&2.0&$0.14$&0.15\nl
&11.5&&&2.0&&&1.86&4.49&2.77&3.3&$0.13$&0.13\nl
&12.0&&&2.5&&&1.77&4.61&2.81&4.5&$0.12$&0.13\nl
&13.0&&&3.5&&&1.65&4.76&2.85&6.9&$0.09$&0.13\nl
&14.0&&&4.5&&&1.57&4.90&2.90&8.2&$0.07$&0.13\nl
&15.0&&&5.5&&&1.51&5.00&2.93&9.7&$0.07$&0.14\nl
DS&10.0&$-0.22$&0.08&0.5&0.00&0.00&4.13&2.40&1.98&0.7&$0.20$&0.17\nl
&10.5&&&1.0&&&2.69&3.20&2.36&1.5&$-0.03$&0.11\nl
&11.0&&&1.5&&&2.20&3.59&2.48&2.7&$-0.07$&0.07\nl
&11.5&&&2.0&&&1.96&3.86&2.59&4.3&$-0.06$&0.06\nl
&12.0&&&2.5&&&1.88&3.93&2.62&5.3&$-0.06$&0.06\nl
&13.0&&&3.5&&&1.77&4.02&2.65&7.6&$-0.10$&0.05\nl
&14.0&&&4.5&&&1.70&4.11&2.68&9.4&$-0.13$&0.05\nl
&15.0&&&5.5&&&1.64&4.19&2.71&10.7&$-0.14$&0.05\nl
GR&10.0&$0.10$&0.21&0.5&0.50&0.25&5.04&1.92&1.55&0.7&$0.59$&0.40\nl
&10.5&&&1.0&&&3.44&2.81&2.13&1.6&$0.41$&0.25\nl
&11.0&&&1.5&&&2.60&3.31&2.43&2.8&$0.32$&0.24\nl
&11.5&&&2.0&&&2.21&3.62&2.58&4.8&$0.25$&0.22\nl
&12.0&&&2.5&&&2.06&3.78&2.64&5.7&$0.25$&0.22\nl
&13.0&&&3.5&&&1.87&4.00&2.70&7.8&$0.23$&0.21\nl
&14.0&&&4.5&&&1.78&4.14&2.73&9.2&$0.23$&0.22\nl
&15.0&&&5.5&&&1.69&4.26&2.76&10.4&$0.23$&0.22\nl
DR&10.0&$-0.22$&0.08&0.5&0.50&0.25&4.95&1.75&1.50&0.8&$0.41$&0.27\nl
&10.5&&&1.0&&&3.42&2.51&2.03&1.9&$0.13$&0.15\nl
&11.0&&&1.5&&&2.62&2.92&2.31&4.0&$0.05$&0.15\nl
&11.5&&&2.0&&&2.27&3.18&2.44&6.5&$0.04$&0.14\nl
&12.0&&&2.5&&&2.13&3.30&2.49&7.6&$0.01$&0.13\nl
&13.0&&&3.5&&&1.96&3.46&2.54&9.5&$-0.01$&0.12\nl
&14.0&&&4.5&&&1.87&3.54&2.55&11.1&$-0.02$&0.12\nl
&15.0&&&5.5&&&1.79&3.62&2.57&12.1&$-0.02$&0.12\nl
\enddata
\tablenotetext{a}{``G'' is the giant elliptical base model
($\sigma=250\;\kms$); ``D'' is the dwarf elliptical base model
($\sigma=100\;\kms$); ``S'' is the solar metallicity, solar
enhancement ratio frosting; ``R'' is the metal-rich, super-solar
enhancement ratio frosting.  All frostings represent 20\% of total
mass.  See text and Figure~\ref{fig:burst_evolve} for details.}
\end{deluxetable}

Four illustrative frosting models are shown in
Figure~\ref{fig:burst_evolve} and Table~\ref{tbl:burst_evolve}.  We
begin by choosing two base populations (lower right) that would fall
on the metallicity hyperplane at age 15 Gyr if they were pure SSPs,
one at $250\;\kms$ (the ``giant'' model) and one at $100\;\kms$ (the
``dwarf'' model).  At an age of 9.5 Gyr in each model, we turn on a
frosting population of 20\% by mass and allow the composite population
to age for a further 5.5 Gyr after this burst, which we identify as
the present time.  Two frosting populations are employed, a
solar-composition model with solar abundance ratios, and a metal-rich
model with $\z=0.5$ and $\enh=0.25$.  Each frosting is combined with
each base, making four models in all.

The evolution of these populations is shown in
Figure~\ref{fig:burst_evolve}.  Initially the composite populations
jump to very young SSP-equivalent ages, moderate-to-high
metallicities, and relatively high \enh\ (long arrows to upper left of
diagram).  As the populations age, the SSP-equivalent ages become
rapidly older while \z\ and \enh\ decrease.  Finally, after several
Gyr, the populations have drifted back close to their starting points,
executing a large loop.

In order to match the data, this scenario must place galaxies back on
the \zp-plane at the present time.  Since $\sigma$ does not change in
this simple model, this means that galaxies must come back to the
correct $\sigma$ contour, allowing for the fact that some galaxies of
their type may not have suffered a star burst and thus continued to
evolve passively to the right.  These evolved points are shown by the
large dots; their corresponding $\sigma$ contours are the two grey
bands, each $\pm1\sigma_{\z}$ wide, where $\sigma_{\z}$ is the rms
residual of \z\ about the plane, i.e., 0.09 dex
(Sec.~\ref{sec:results_tzs}).  If a frosting galaxy winds up in the
appropriate grey band after 5.5 Gyr, we will count it as lying in the
\zp-plane, and the model is a success.

Whether or not this will happen depends on a proper match between the
metallicity of the base population and that of the frosting.  The
giant base ($\z\ = +0.1$) plus metal-rich frosting ($\z\ = +0.5$) is
an example of a successful combination; it falls exactly in the middle
of the allowed grey band at the present time (upper solid model).  The
same base enriched with a \emph{solar}-metallicity frosting is less
successful because the combination falls \emph{below} the allowed grey
band (upper dotted model).  From these two models, it can be seen that
the metal abundance of a successful frosting must be between 0.1 and
0.6 dex more metal rich than the giant base population to which it is
added.  Similar reasoning implies that the same window---0.1--0.6 dex
more metal rich---applies to dwarf bases, too.

The width of these windows depends on the age of the starburst.
Turning on the starburst 5.5 Gyr ago was arbitrary and resulted in
fairly red, old-looking models at the present time.  Since many SSPs
are observed to be quite young, matching them requires more recent
starbursts.  Metallicity constraints then get tighter---it may be
shown that the allowed \z\ window shrinks in width and the frosting
population must be considerably \emph{more} metal-rich than the base.

Apparent \enh\ values must also stay constant during this process,
since by hypothesis $\sigma$ is assumed not to change
(Eq.~\ref{eq:sa}; Fig.~\ref{fig:sa}).  This further requires that
\enh\ for the frosting population be nearly equal to that of the base
population, as composite SSP enhancement ratio is close to the mean of
the frosting and base populations at moderate burst strength (this
point was also made by J{\o}rgensen 1999).

The close coordination required for both \z\ and \enh\ in frosting
models may place tight constraints on star formation scenarios for
elliptical galaxies.  In particular, it seems hard to meet the
necessary tight limits on \z\ and \enh\ if young populations form from
unrelated, ``foreign'' gas acquired in a merger.  Such coordination
would seem more natural if the frosting gas were pre-enriched
\emph{within the parent galaxy itself}.  An example of such a model
might be low-mass star formation in gas re-accreted in a galactic
cooling flow (Mathews \& Brighenti 1999).

Several questions remain about the frosting scenario:

(1) The frosting model as presented here consists of only two bursts.
More realistic scenarios would contain extended star formation over
time.

(2) Some of the most extreme young populations in the present sample
are clearly in disturbed galaxies: NGC 6702 (\cite{Davoust87}; Tonry
et al., priv.~comm.), NGC 1700, NGC 584, and NGC 5831 (\cite{SS92}),
which are excellent candidates for recent star formation in mergers.
We have argued that such star formation would likely disobey the
hyperplane, yet these objects fall nicely on it
(Figures~\ref{fig:pca}, \ref{fig:tz}).  Their agreement with the
hyperplane suggests that the previous argument against foreign gas
captured in mergers may not be fully correct.

(3) The stellar population parameters considered here are only central
values (\reo{8}).  The global stellar populations (\reo{2}) are
generally older by $\sim25\%$ and more metal-poor by $-0.20$ dex,
while \enh\ is basically the same (Paper I).  We believe that global
populations also obey a hyperplane but have not yet examined it in
detail.  Radial gradients and global stellar populations will be
discussed in a future paper.

\section{Conclusions}\label{sec:conc}

The centers of local elliptical galaxies appear to contain quite
complex stellar populations.  The present sample of local ellipticals
spans a wide range of stellar population parameters, most notably a
large range in SSP-equivalent age (especially in, but not limited to,
field ellipticals).

Despite their diversity, the central stellar populations of these
galaxies are described by a few simple scaling relations.  (1)
Abundance parameters \z\ and \enh\ are specified to high accuracy by
SSP-equivalent age, $t$, and central velocity dispersion $\sigma$;
ellipticals thus occupy a ``metallicity hyperplane'' in
$(\z,\logt,\log\sigma,\enh)$-space.  (2) SSP-equivalent metallicity,
\z, is a function of both $t$ and $\sigma$ (the ``\zp-plane'').  At
fixed $t$, \z\ increases with $\sigma$; at fixed $\sigma$, \z\ is
larger at younger age.  (3) SSP-equivalent enhancement ratio, \enh, is
found to be a monotonically increasing function of $\sigma$ only, in
the sense that adding other structural parameters such as $I_e$ or
$r_e$ does not predict either \enh\ or \z\ more accurately.

Our use of SSP-equivalent parameters is not meant to imply a
single-burst origin for elliptical galaxies; in fact, the existence of
the \zp-plane seems to imply that the populations are largely old with
a ``frosting'' extending to younger ages.  However, despite the fact
that SSP-parameters are not true means but rather likely to be
influenced by the light of younger stars, they still place very
important constraints on the history of star formation in elliptical
galaxies (see below).

We take $\sigma$ and $t$ as the independent parameters that specify
the distribution of galaxies in the hyperplane.  Any variation in this
distribution will influence all other two-dimensional projections of
SSP parameters, and thus many of the common scaling laws for
elliptical galaxies.  Our sample shows a possible difference in the
($\sigma, t$) distribution with environment---our field ellipticals
span a wide range in SSP age, while the Fornax and Virgo ellipticals
are generally old.  This results in a significant mass-metallicity
trend for the cluster galaxies but not for the field galaxies.  Other
correlations between stellar population parameters and structural
parameters may also turn out to vary with environment.

The Mg--$\sigma$ relations are edge-on projections of the metallicity
hyperplane.  At a given $\sigma$, young age is offset by a
correspondingly high metallicity, preserving line strength.  The
narrowness of the observed Mg--$\sigma$ relations therefore does not
imply a narrow range of ages at fixed velocity dispersion.  A more
detailed look at the Mg--$\sigma$ relations is the subject of a future
paper.

Physical models to account for the hyperplane have been considered.
The rise in \enh\ with $\sigma$ and the mass-metallicity relation (at
fixed $t$) is consistent with a higher effective yield of Type II SNe
products at high $\sigma$.  This trend has several possible
explanations, for example, greater retention of outflow-driven gas or
a flatter IMF at high $\sigma$.

The existence of the \zp-plane is more challenging.  A ``frosting''
scenario is favored, in which young stars are added to an old base
population, resulting in a range of SSP-equivalent ages.  With a
suitable choice of burst populations, the composite populations can be
engineered to lie on lines of constant $\sigma$ in the \zp-plane after
a few Gyr.  However, to preserve both the \zp-plane and the
\enh--$\sigma$ relation requires that abundances in the frosting
population must be closely coupled to that of the base
population---the metallicity, \z, of the frosting must be somewhat
higher than that of the base population, while the enhancement ratio,
\enh, must be nearly equal.  The frosting scenario therefore seems to
favor star formation from gas that was pre-enriched in the same parent
galaxy rather than from gas that was accreted in an unrelated merger.
However, several merger remnants in the sample do indeed lie nicely on
the \zp-plane, in defiance of this logic.

The present picture of the hyperplane is preliminary and needs to be
checked against a better local sample and a wide array of other data.
For example, SSP mass-to-light ratios should be compared to dynamical
$M/L$ measurements, and \emph{global} SSP parameters should be
analyzed, as they are more indicative of the global star formation
history than the central SSP parameters used here.  A further
interesting question is whether the color-magnitude relation and other
scaling laws might also be near-edge-on projections of the hyperplane,
like Mg--$\sigma$.

Most important, the implications of the frosting model must be
developed for lookback observations of distant elliptical galaxies.
Many observations of distant cluster ellipticals suggest that their
stellar populations formed very early, and this may be consistent with
the generally old ages for cluster galaxies found here.  Our field
ellipticals do show a wide spread of SSP ages, but we have noted that
the sample is not volume-limited, and thus predictions for the
evolution of distant field ellipticals cannot yet be drawn.  In short,
a great deal more data must be gathered and reconciled before we can
claim a solid understanding of the star formation histories of
elliptical galaxies.

\acknowledgments

It is a pleasure to thank a great number of our colleagues for
interesting discussions.  Drs.~R. Bender, M. Bolte, D. Burstein,
R. Carlberg, J. Dalcanton, R. Davies, A. Dressler, R, Ellis,
W. Freedman, G. Illingworth, D. Kelson, I. King, R. Marzke,
W. Mathews, A. McWilliam, J. Mould, J. Mulchaey, A. Oemler,
A. Renzini, M. Rich, P. Schechter, F. Schweizer, T. Smecker-Hane,
P. Stetson, S. Yi, and A. Zabludoff have all provided hours of
stimulating conversations.  We are indebted to Dr.~M. Tripicco for
sending us electronic versions of his and Dr.~Bell's results on the
response of the Lick/IDS indices to abundance variations, to
Dr.~H. Kuntschner for providing his data on Fornax early-type galaxies
in advance of publication, to Dr. D.~Kelson for plane-fitting
software, and to Drs.~J. Tonry, J. Blakeslee, and A. Dressler for
providing SBF distances to local ellipticals in advance of publication
and for allowing S. C. T. to examine their images of NGC 6702.  The
comments of an anonymous referee helped greatly to improve the
presentation.  Support for this work was provided by NASA through
Hubble Fellowship grant HF-01125.01-99A to SCT awarded by the Space
Telescope Science Institute, which is operated by the Association of
Universities for Research in Astronomy, Inc., for NASA under contract
NAS 5-26555, by a Starr Fellowship to SCT, by a Flintridge Foundation
Fellowship to SCT, and by NSF grant AST-9529098 to SMF.

\appendix

\section{Models of composite stellar populations}\label{app:comp}

In this section we discuss simple models of composite stellar
populations based on double bursts.  Our approach is similar to the
``isochrone synthesis'' method of Bruzual \& Charlot (1993), in which
composite populations are built up from single stellar populations
(SSPs) treated as $\delta$-functions.

At present, it is not our intent to create grids of models with
multiple populations drawn from galaxy formation and evolution models
including the effects of winds, blowout, and other processes (see
\cite{AY87} and \cite{TCBMP98} for two examples of this approach).
Rather, we are interested in determining rough rules of thumb for
adding multiple populations in the \hbeta-metallicity diagrams.
Specifically we ask how mixtures of two bursts or multiple
metallicities combine to mimic a single SSP of a given age and
metallicity.

We begin by describing the method used to combine the W94 SSPs to
derive line strengths.  We then discuss two models of composite
populations: galaxies with multiple (here, two) bursts of star
formation, and a model with a single age but a dispersion in
metallicity based on the metallicity spread of M32 as determined by
Grillmair et al.\ (1996).  We show that line strengths add as vectors
in the diagrams to first order (when weighted by light).  There is
thus an infinite number of ways of decomposing a given population into
single-burst components.  Determining the detailed star formation
histories of old stellar populations from the present data is highly
underconstrained.

\subsection{Method}

The line strength for a single stellar population (when expressed as
an equivalent width in \AA) can be written
\begin{equation}
\mathrm{EW}=w\left(1-\frac{F_I}{F_C}\right),
\end{equation}
where $w$ is the width of the feature bandpass in \AA, $F_I$ is the
observed flux (per unit mass) integrated over the feature bandpass,
and $F_C$ represents the observed flux (per unit mass) of the straight
line connecting the midpoints of the blue and red pseudocontinuum
levels, integrated over the feature bandpass (\cite{WFGB94};
\cite{TWFBG98}; Paper I).  In a composite population, the fluxes
become sums over populations and therefore
\begin{equation}
\mathrm{EW}=w\left(1-\frac{\sum_i f_i F_{I,i}}{\sum_i f_i F_{C,i}}\right),
\label{eq:cspew}
\end{equation}
where $i$ represents each individual population, $f_i$ is the fraction
by mass of each population ($\sum_i f_i=1$), and $F_{I,i}$ and
$F_{C,i}$ are the integrated fluxes in the feature bandpass and in the
``continuum'' of each population $i$.

We assume that $F_{C,i}$ is independent of $\enh\neq0$, which is
consistent with the tracks of Salaris \& Weiss (1998), in which the
turnoff and RGB move horizontally but do not change luminosity.  For
each population we can then write
\begin{equation}
F_{I,i} = F_{C,i}\left(1 - \frac{I_i(t,\z,\enh)}{w}\right),
\label{eq:enhfline}
\end{equation}
where $I_i(t,\z,\enh)$ is the line strength of population $i$ for the
index in question at age $t$, metallicity \z, and enhancement ratio
\enh.  The model values $F_{I,i}$ and $F_{C,i}$ values are then
inserted into Equation~\ref{eq:cspew} to determine the line strength
of the composite population for each index of interest.

\subsection{Models}

\subsubsection{Double starbursts}

Three double-starburst models are developed, chosen illustratively
such that their composite line strengths cover the observed loci of
the G93 galaxies.  Model A covers giant ellipticals with
$\sigma\ga200\;\kms$; its old component has ($t$,\z,\enh) = (17 Gyr,
+0.15, +0.25), similar to the oldest galaxies in the sample, mixed
with a young burst having parameters (1 Gyr, +0.75, 0.0).  Model B
covers small ellipticals with $\sigma\la200\;\kms$.  Its old
population has (17 Gyr, $-0.25$, 0.0), mixed with a young population
of (1 Gyr, +0.5, 0.0).  Model C is an alternative to model B in which
metal-enriched winds are imagined to selectively blow out SN II
products but not those from SN Ia (\cite{MF98}).  Its old population
has (17 Gyr, $-$0.25, +0.25), and its young burst has (1 Gyr, +0.5,
$-$0.25) (highly enriched in SN Ia products).  In all models, the
young burst is allowed to vary in strength from 10\%--100\% of the
final mass.

\begin{figure*}
\setcounter{figure}{0}
\plotone{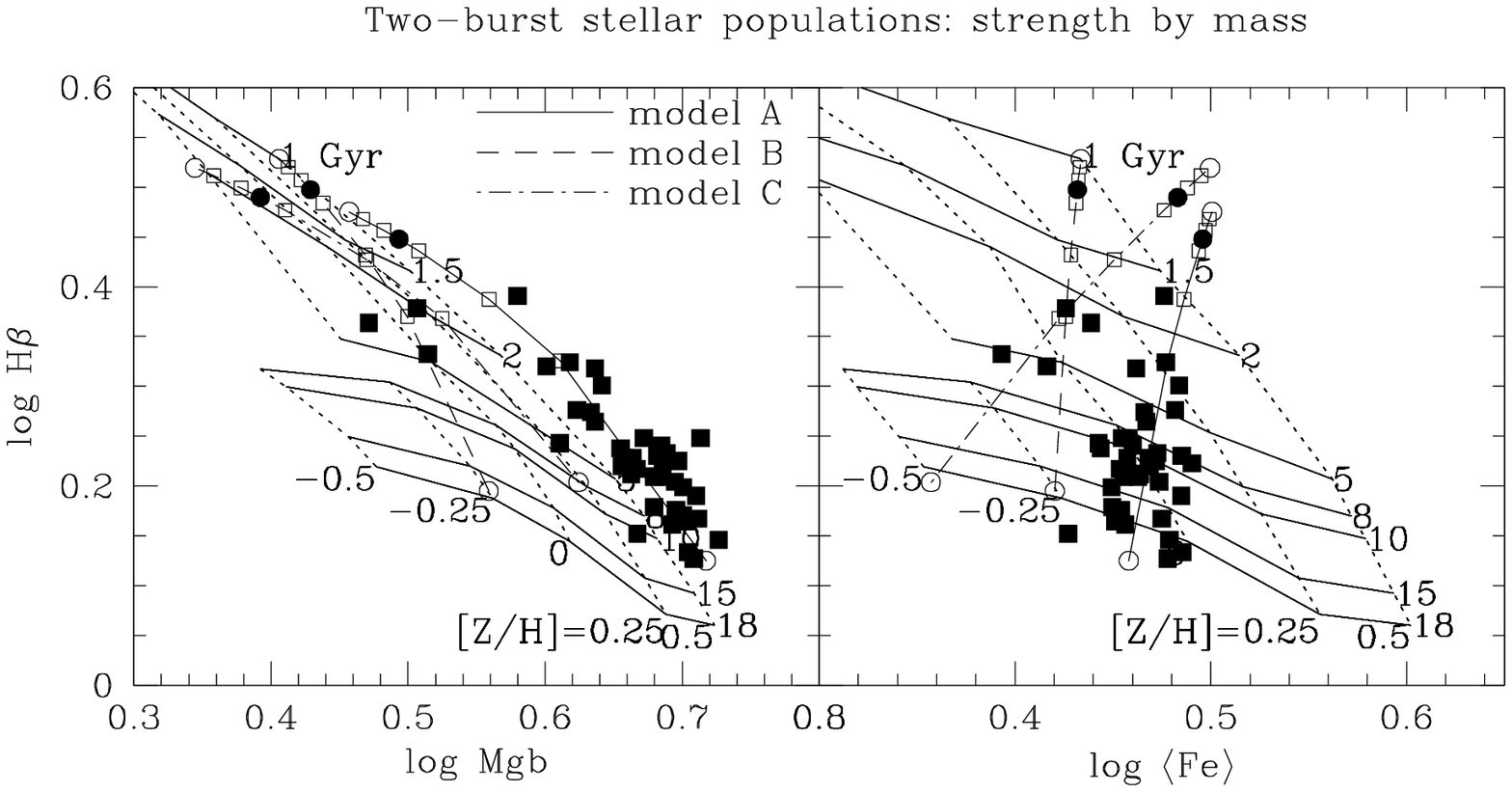}
\caption{Schematic two-burst models.  Three models are shown: (A) a 17
Gyr, $\z=+0.15$ dex, $\enh=+0.25$ dex progenitor (typical of the
oldest giant ellipticals in the sample) with a 1 Gyr, $\z=+0.75$ dex,
solar-neighborhood abundance ratio burst, meant to cover the stellar
populations of the high-$\sigma$ galaxies (solid line); (B) a 17 Gyr,
$\z=-0.25$ dex, solar-neighborhood abundance ratio progenitor with a 1
Gyr, $\z=+0.5$ dex, solar-neighborhood abundance ratio burst, meant to
cover the stellar populations of the low-$\sigma$ galaxies NGC 221
(M32), NGC 4489, and NGC 7454 (short-dashed line); and (C) a 17 Gyr,
$\z=-0.25$ dex, $\enh=+0.25$ dex progenitor with a 1 Gyr, $\z=+0.5$
dex, $\enh=-0.25$ burst dex, meant to represent possible star
formation after a metal-enriched wind in a low-$\sigma$ galaxy
(dot-dashed line).  Bursts of 10\%, 20\%, 40\%, 60\%, and 80\% (open
squares) and 50\% (solid circles) by mass are shown.  Open circles
represent the progenitor (lower) and burst (upper) populations.  Solid
squares are the G93 galaxies; compare to
Figure~\ref{fig:re8m}.\label{fig:twoburst}}
\end{figure*}

\begin{deluxetable}{ccccccccccccccccc}
\tablefontsize{\footnotesize}
\setcounter{table}{0}
\tablecaption{Two-burst composite stellar population
models\label{tbl:twoburst}}
\tablewidth{0pt}
\tablehead{
&\multicolumn{4}{c}{Base}&
\multicolumn{4}{c}{Frosting}&
\multicolumn{8}{c}{Composite} \\
\colhead{Model}&\colhead{$t$}&\colhead{\z}&\colhead{\enh}&\colhead{$M/L_V$}&
\colhead{$t$}&\colhead{\z}&\colhead{\enh}&\colhead{$M/L_V$}&
\colhead{$f_M$\tablenotemark{a}}&
\colhead{$f_L$\tablenotemark{b}}&
\colhead{\hbeta}&\colhead{\mgb}&\colhead{\fe}&
\colhead{$t$}&\colhead{\z}&\colhead{\enh}}
\tablecolumns{20}
\startdata
A&17&$+0.15$&$+0.25$&10.0&1&$+0.75$&\phs0.00&1.3&0.10&0.46&2.12&4.08&2.88&2.2&$+0.50$&$+0.15$\nl
&&&&&&&&&0.12&0.50&2.19&3.98&3.01&2.0&$+0.49$&$+0.13$\nl
&&&&&&&&&0.40&0.84&2.73&3.22&3.12&1.3&$+0.69$&$+0.03$\nl
B&17&$-0.25$&\phs0.00&\phn7.9&1&$+0.50$&\phs0.00&1.2&0.10&0.42&2.35&3.16&2.67&2.8&$+0.08$&\phs$0.00$\nl
&&&&&&&&&0.14&0.50&2.50&3.07&2.67&2.1&$+0.15$&\phs$0.00$\nl
&&&&&&&&&0.40&0.81&3.05&2.74&2.70&1.3&$+0.38$&\phs$0.00$\nl
C&17&$-0.25$&$+0.25$&\phn7.9&1&$+0.50$&$-0.25$&1.2&0.10&0.42&2.33&3.35&2.65&2.7&$+0.14$&$+0.06$\nl
&&&&&&&&&0.14&0.50&2.48&3.18&2.72&1.9&$+0.23$&$+0.02$\nl
&&&&&&&&&0.40&0.81&3.00&2.57&2.99&1.3&$+0.42$&$-0.14$\nl
\enddata
\tablenotetext{a}{Fractional mass of burst}
\tablenotetext{b}{Fraction of $V$-band light in burst}
\end{deluxetable}

\begin{figure*}
\plotone{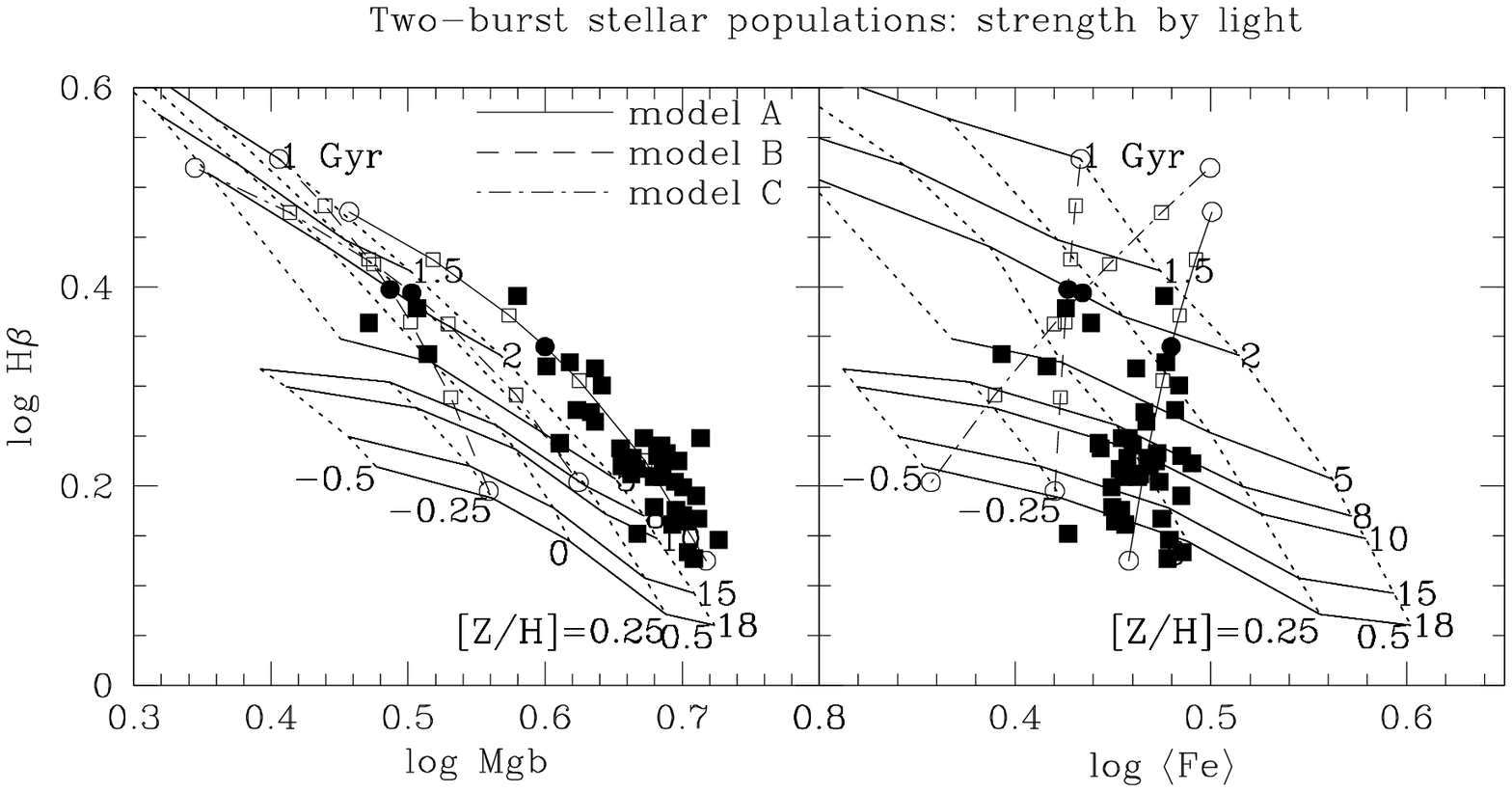}
\caption{The same two-burst models as in Figure~\ref{fig:twoburst},
but with burst strength now indicated by light fraction.  Open symbols
represent the fractional V-band luminosity of the young population:
from bottom to top, 0\%, 20\%, 40\%, 60\%, 80\% and 100\%.  These
correspond to average mass fractions of 0\%, 3\%, 8\%, 16\%, 35\%, and
100\% for model A and 0\%, 4\%, 9\%, 19\%, 38\%, and 100\% for models
B and C.  The solid circles represent the 50\% by light in young/old
models, which correspond to 12\% by mass in the young population for
model A and 14\% by mass in the young population for models B and C.
The relative straightness of the lines and even spacing of the squares
compared to Figure~\ref{fig:twoburst} indicate that stellar
populations add roughly as light-weighted vectors in these diagrams.
\label{fig:twoburstl}}
\end{figure*}

The models are summarized in Table~\ref{tbl:twoburst} and illustrated
in Figure~\ref{fig:twoburst}, which shows the weights expressed as
percentage of \emph{mass}, and in Figure~\ref{fig:twoburstl}, which
shows the weights expressed as percentage of \emph{light}.  The latter
figure demonstrates the useful rule of thumb that composite, two-burst
populations add roughly as light-weighted vectors in the Balmer--metal
line strength diagrams.  This is shown by the relatively straight
lines linking the endpoint populations in Figure~\ref{fig:twoburstl}
and the relatively uniform tickmark spacing along the lines.  Taking
model C as an example, we can compare the light-weighted vector rule
for predicting the 50/50 population, versus its actual location in the
diagrams.  For model C, the real composite 50/50 population (50\% old,
50\% young by light) is at (1.9 Gyr, $+0.23$, $+0.02$) while the
vector-added point midway between the two endpoints is at (2.5 Gyr,
0.0, $-0.05$).  For an 80/20 model (80\% old, 20\% young by light),
the real population is at (7.3 Gyr, $-0.04$, +0.15) compared to the
vector-added population at (9.5 Gyr, $-0.15$, +0.15).  Thus, vector
weighting by light tends to overestimate the age by about 25\%,
underestimate \z\ by 0.1--0.25, and underestimate \enh\ by less than
about 0.1.  These are extreme cases, and the errors for mixing two
populations closer in the diagrams would be smaller.

In the past, we have stated that the best-fitting SSP-equivalent age
(as derived here) is close to the ``light-weighted'' age
(\cite{FTGW95}).  This was a mis-statement.  The light-weighted age of
the 50/50 model is simply the average of 1 Gyr and 17 Gyr, or 9 Gyr,
much larger than the SSP-equivalent age, of 1.8 Gyr.  What we meant to
say is that composite populations add in the diagrams \emph{like
light-weighted vectors}.  As noted, the age agreement is much better,
within 50\%, when computed this way.  However, valuable as such rules
of thumb may be for cultivating intuition, the only proper way to
compare models to data is to add up the fractional index contributions
using Equation~\ref{eq:cspew}.

The light-weighted vector rule cannot be taken too far and does better
for \fe\ than for \mgb, whose trajectories are not as straight in the
grid diagrams.  This may prove to be a boon in accounting for the very
high \mgb\ strengths of galaxies like NGC 507, NGC 6702, and NGC 720,
whose \mgb\ indices lie high up and to the right in
Figure~\ref{fig:re8m}.  Such populations might be modeled as recent
starbursts, as suggested independently by their high morphological
disturbance parameters (\cite{FTGW95}).

\subsubsection{Metallicity spreads}

Yet a fourth model (not shown) explores the effect of a spread in
metallicities at a single age.  This model is based on the metallicity
distribution in an outer field of M32 determined by Grillmair et al.\
(1996; their Figure~10), which has a strong peak at $\feh=\z=-0.20$,
FWHM of about 0.5, a weak tail to low metallicities down to $-$1.2,
and a light-weighted mean metallicity of $-0.25$ (note that
$\enh\approx0.00$ for M32).  For an assumed single age of 8.5 Gyr, the
composite model yields $\hbeta=2.02$ \AA, $\mgb=2.89$ \AA, and
$\fe=2.29$ \AA, in good agreement with the outwardly extrapolated data
from G93 of $\hbeta=1.92$ \AA, $\mgb=2.99$ \AA, and $\fe=2.42$ \AA\
(Grillmair et al.~1996).

The SSP-equivalent stellar population parameters of the composite
model are $t=8.2$ Gyr, $\z=-0.32$, and $\enh=0.00$.  These results
show that the integrated light from a uniform-age population with a
strongly peaked metallicity distribution resembles a population of
nearly the same age (or slightly younger if metal-poor stars are
present) and of very similar \z\ to the true light-weighted
metallicity ($\z=-0.25$).  These results agree with composite
multi-metallicity populations by Greggio (1997), who found that shifts
of SSP-equivalent metallicities in mixed-metallicity populations were
not large in the absence of large metal-poor tails.

\bigskip
To summarize, the results in this Appendix suggest that metallicity
spreads (and, by extension, spreads in \enh) at fixed age do not
seriously skew the indices, but that even small populations of
recently-formed (within $\sim1$ Gyr) stars can significantly reduce
the inferred age.  A burst of only 10\% by mass 1 Gyr ago on top of a
17 Gyr old population gives an SSP-equivalent age of only $\approx1.8$
Gyr.  Because line strengths add as vectors (weighted by the
luminosity of each population), the ages and metallicities of each
burst in a composite population are not separable using the present
data.

\end{document}